\documentclass[aip,reprint]{revtex4-1}
\usepackage{graphicx}
\usepackage{hyperref}
\usepackage{xcolor}
\usepackage{fancyhdr}

\fancypagestyle{firststyle}
{
\fancyhf{}

\rfoot{\thepage}
\pagestyle{fancy}
}

\usepackage{fancyhdr}

\begin{document}

\title{\textcolor{blue}{Probabilistic computing with p-bits}} 

\author{Jan Kaiser}
\email[]{kaiser32@purdue.edu}

\affiliation{Elmore Family School of Electrical and Computer Engineering, Purdue University, West Lafayette, IN, 47906 USA}

\author{Supriyo Datta }
\affiliation{Elmore Family School of Electrical and Computer Engineering, Purdue University, West Lafayette, IN, 47906 USA}

\date{\today}

\begin{abstract}
Digital computers store information in the form of $bits$ that can take on one of two values $0$ and $1$, while quantum computers are based on $qubits$ that are described by a complex wavefunction whose squared magnitude gives the probability of measuring either a $0$ or a $1$. Here we make the case for a probabilistic computer based on \textit{p-bits} which take on values $0$ and $1$ with controlled probabilities and can be implemented with specialized compact energy-efficient hardware. We propose a generic architecture for such \textit{p-computers} and emulate systems with thousands of p-bits to show that they can significantly accelerate randomized algorithms used in a wide variety of applications including but not limited to Bayesian networks, optimization, Ising models and quantum Monte Carlo.
\end{abstract}


\maketitle 
\thispagestyle{firststyle}

\vspace{-0.25in}

\section{Introduction}
Feynman \cite{feynman_simulating_1982} famously remarked \textit{``Nature isn't classical, dammit, and if you want to make a simulation of nature, you'd better make it quantum mechanical"}. In the same spirit we could say \textit{``Many real life problems are not deterministic, and if you want to simulate them, you'd better make it probabilistic"}. But there is a difference. Quantum algorithms require quantum hardware and this has motivated a worldwide effort to develop a new appropriate technology. By contrast probabilistic algorithms can be and are implemented on existing deterministic hardware using \textit{pseudo RNG's} (random number generators). Monte Carlo algorithms represent one of the top ten algorithms of the $20^{th}$ century \cite{cipra_best_2000} and are used in a broad range of problems including Bayesian learning, protein folding, optimization, stock option pricing, cryptography just to name a few. So why do we need a \textit{p-computer}?

\begin{figure*}
    \setlength\abovecaptionskip{-0.5\baselineskip}
    \centering
   \includegraphics[width=0.9\linewidth]{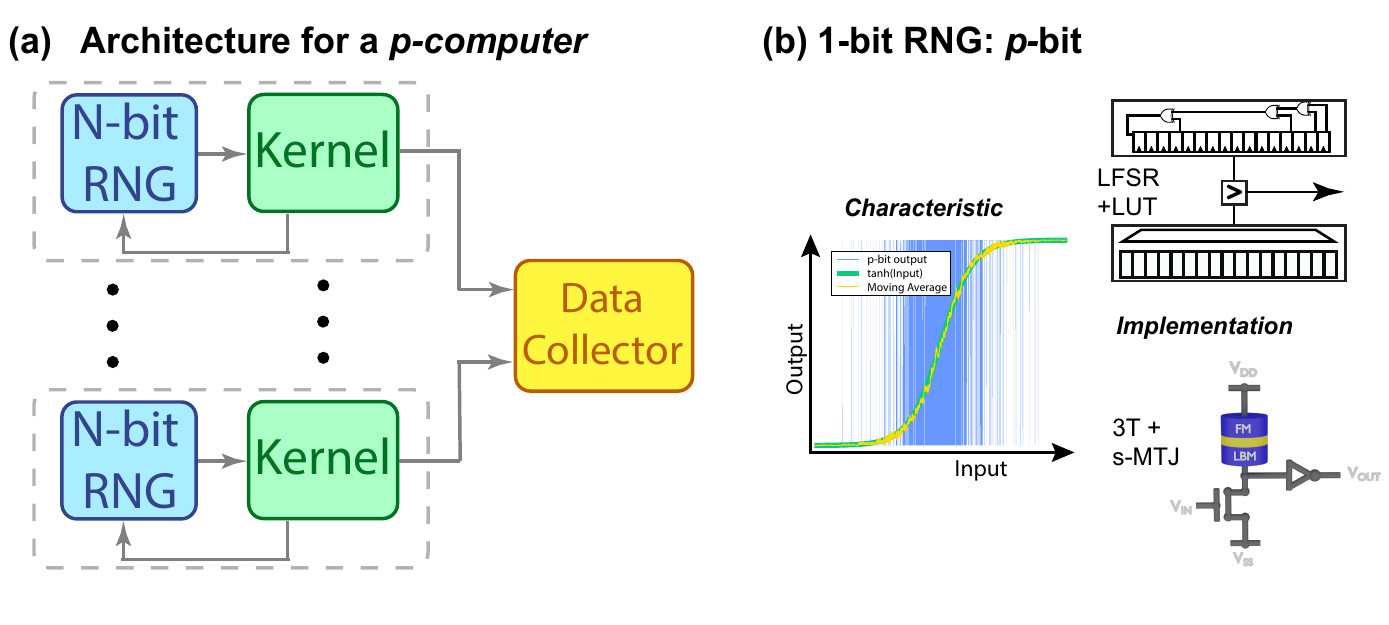} 
    \vspace{0in}
        \caption{\textbf{Probabilistic computer: (a)} Overall architecture combining a probabilistic element (N-bit RNG) with deterministic elements (kernel and data collector). The N-bit RNG block is a collection of N 1-bit RNG's, or p-bits. \textbf{(b)} p-bit: Desired input-output characteristic along with two possible implementations, one with CMOS technology using linear feedback shift registers (LFSR's) and lookup tables (LUT's)\cite{sutton_autonomous_2020} and the other using three transistors and a stochastic magnetic tunnel junction (s-MTJ)\cite{camsari_implementing_2017}. The first is used to obtain all the results presented here, while the second is a nascent technology with many unanswered questions. }
     \label{fig: fig1}
\end{figure*}

\begin{figure*}
    \setlength\abovecaptionskip{-0.5\baselineskip}
    \centering
    \includegraphics[width=0.85\linewidth]{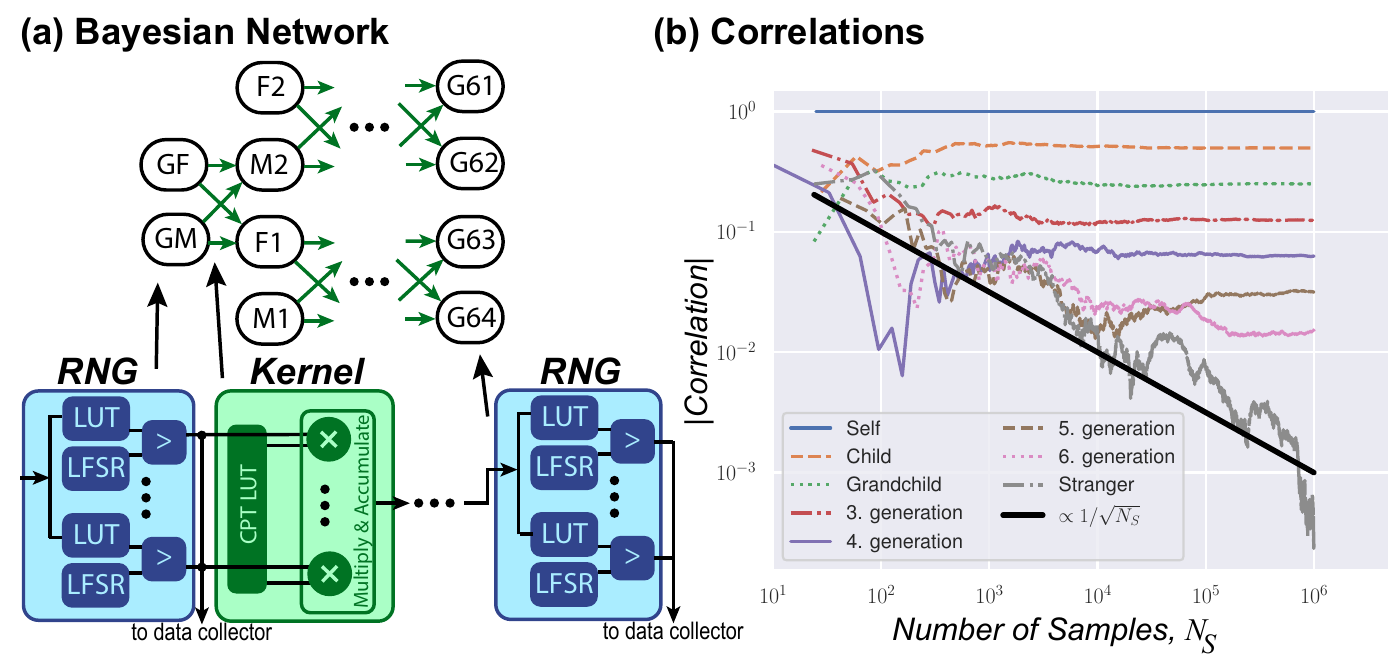} 
    \vspace{0.2in}
    \caption{\textbf{Bayesian network} for genetic relatedness mapped to a \textit{p-computer} \textbf{(a)} with each node represented by one p-bit. With increasing $N_{S}$, the correlations \textbf{(b)} between different nodes is obtained more accurately.}
    \label{fig: fig2}
\end{figure*}

A key element in a Monte Carlo algorithm is the RNG which requires thousands of transistors to implement with deterministic elements, thus encouraging the use of architectures that time share a few RNG's. Our work has shown the possibility of high quality true RNG's using just three transistors\cite{camsari_implementing_2017}, prompting us to explore a different architecture that makes use of large numbers of controlled-RNG's or \textit{p-bits}. Fig. \ref{fig: fig1} (a) \cite{kaiser_benchmarking_2021} shows a generic vision for a probabilistic or a \textit{p-computer} having two primary components: an \textit{N-bit random number generator (RNG)} that generates N-bit samples and a \textit{Kernel} that performs deterministic operations on them. Note that each \textit{RNG-Kernel} unit could include multiple \textit{RNG-Kernel} sub-units (not shown) for problems that can benefit from it. These sub-units could be connected in series as in Bayesian networks (Fig. \ref{fig: fig2} (a) or in parallel as done in parallel tempering\cite{swendsen_replica_1986-1,earl_parallel_2005} or for problems that allow graph coloring\cite{ko_accelerating_2019}. The parallel RNG-Kernel units shown in Fig. \ref{fig: fig1} (a) are intended to perform easily parallelizable operations like ensemble sums using a \textit{data collector} unit to combine all outputs into a single consolidated output. 

Ideally the \textit{Kernel} and \textit{data collector} are pipelined so that they can continually accept new random numbers from the $RNG$ \cite{kaiser_benchmarking_2021}, which is assumed to be fast and available in numerous numbers. The \textit{p-computer} can then provide $N_p f_c$ samples per second, $N_p$ being the number of parallel units\footnote{Here, we assume that every RNG-Kernel unit gives one sample per clock cycle. There could be cases where multiple samples could be extracted from one unit per clock cycle.}, and $f_c$ the clock frequency. We argue that even with $N_p=1$, this throughput is well in excess of what is achieved with standard implementations on either \textit{CPU} (central processing unit) or \textit{GPU} (graphics processing unit) for a broad range of applications and algorithms including but not limited to those targeted by modern \textit{digital annealers} or \textit{Ising solvers} \cite{goto_combinatorial_2019,aramon_physics-inspired_2019,yamamoto_73_2020,yamaoka_243_2015,ahmed_probabilistic_2020,sutton_autonomous_2020,patel_ising_2020,cai_power-efficient_2020,dutta_ising_2021}.
Interestingly, a \textit{p-computer} also provides a conceptual bridge to quantum computing, sharing many characteristics that we associate with the latter\cite{camsari_dialogue_2021}. Indeed it can implement algorithms intended for quantum computers, though the effectiveness of \textit{quantum Monte Carlo} depends strongly on the extent of the so-called sign problem specific to the algorithm and our ability to `tame' it \cite{troyer_computational_2005}.

\section{Implementation}

Of the three elements in Fig. \ref{fig: fig1}, two are \textit{deterministic}. The \textit{Kernel} is problem-specific ranging from simple operations like addition or multiplication to more elaborate operations that could justify special purpose chiplets \cite{li_chiplet_2020}. Matrix multiplication for example could be implemented using analog options like resistive crossbars \cite{liu_spiking_2015,li_memristor-based_2013,cai_power-efficient_2020,demler_mythic_2018}. The data collector typically involves addition and could be implemented with adder trees. The third element is \textit{probabilistic}, namely the N-bit \textit{RNG} which is a collection of $N$ 1-bit RNG's or $p \textendash bits$. The behavior of each $p \textendash bit$ can be described by \citep{camsari_stochastic_2017}
\begin{equation}
s_i = \Theta \ [\ \sigma(I_i - r) \ ]
\label{eqn: eqn1}
\end{equation}
where $s_i$ is the binary p-bit output, $\Theta$ is the step function, $\sigma$ is the sigmoid function, $I_i$ is the input to the p-bit and $r$ is a uniform random number between 0 and 1. Eq. (\ref{eqn: eqn1}) is illustrated in Fig. \ref{fig: fig1} (b). While the p-bit output is always binary, the p-bit input $I_i$ influences the mean of the output sequence. With $I_{i} = 0$, the output is distributed $50-50$ between $0$ and $1$ and this may be adequate for many algorithms. But in general a non-zero $I_{i}$ determined by the current sample is necessary to generate desired probability distributions from the \textit{N-bit RNG}-block.

One promising implementation of a p-bit is based on a stochastic magnetic tunnel junction (s-MTJ) as shown in Fig. \ref{fig: fig1} (b) whose resistance state fluctuates due to thermal noise. It is placed in series with a transistor, and the drain voltage is thresholded by an inverter \citep{camsari_implementing_2017} to obtain a random binary output bit whose average value can be tuned through the gate voltage $V_\mathrm{IN}$. It has been shown both theoretically \cite{kaiser_subnanosecond_2019,kanai_theory_2021} and experimentally \cite{safranski_demonstration_2021,hayakawa_nanosecond_2021} that \textit{s-MTJ}-based p-bits can be designed to generate new random numbers in times $\sim$\textit{ nanoseconds}. The same circuit could also be used with other fluctuating resistors\cite{hassan_quantitative_2021}, but one advantage of \textit{s-MTJ's} is that they can be built by modifying magnetoresistive random access memory (MRAM) technology that has already reached gigabit levels of integration \cite{borders_integer_2019}.

Note, however, that the examples presented here all use p-bits implemented with deterministic CMOS elements or \textit{pseudo-RNG's} using linear feedback shift registers (LFSR's) combined with lookup tables (LUT's) and thresholding elements\cite{sutton_autonomous_2020} as shown in Fig. \ref{fig: fig1} (b). Such random numbers are not truly random, but have a period that is longer than the time range of interest. The longer the period, the more registers are needed to implement it. Typically a p-bit requires  $\sim1000$ transistors \cite{borders_integer_2019}, the actual number depending on the quality of the pseudo RNG that is desired. Thirty-two stage LFSR's require $\sim 1200$ transistors, while a Xoshiro128+ \cite{vigna_further_2017} would require around four times as many. Physics-based approaches, like \textit{s-MTJ's}, naturally generate true random numbers with infinite repetition period and ideally require only 3 transistors and 1 MTJ.

A simple performance metric for \textit{p-computers} is the ideal sampling rate $N_p f_c$ mentioned above. The results presented here were all obtained with an field-programmable gate array (FPGA) running on a 125 MHz clock, for which $1/f_c = 8$ ns, which could be significantly shorter (even $\sim 0.1$ ns \cite{kaiser_subnanosecond_2019}) if implemented with \textit{s-MTJ's}. Furthermore, \textit{s-MTJ's} are compact and energy-efficient, allowing up to a factor of 100 larger $N_p$ for a given area and power budget. With an increase of $f_c$ and $N_p$, a performance improvement by \textit{2-3 \ orders} of magnitude over the numbers presented here may be possible with \textit{s-MTJ's} or other physics-based hardware.

We should point out that such compact \textit{p-bit} implementations are still in their infancy \cite{borders_integer_2019} and many questions remain. First is the inevitable variation in RNG characteristics that can be expected. Initial studies suggest that it may be possible to train the $Kernel$ to compensate for at least some of these variations \cite{kaiser_probabilistic_2020,kaiser_hardware-aware_2021}. Second is the quality of randomness, as measured by statistical quality tests which may require additional circuitry as discussed for example in Ref. \cite{safranski_demonstration_2021}. Certain applications like simple integration (Section \ref{subsec: integration}) may not need high quality random numbers, while others like Bayesian correlations (Section \ref{subsec: Bayes}) or Metropolis-Hastings methods that require a proposal distribution (Section \ref{subsec: Knapsack}) may have more stringent requirements. Third is the possible difficulty associated with reading sub-nanosecond fluctuations in the output and communicating them faithfully.  Finally, we note that the input to a p-bit is an analog quantity requiring Digital-to-Analog Converters (DAC's) unless the kernel itself is implemented with analog components.

\section{Applications}

\subsection{Simple integration}
\label{subsec: integration}
A variety of problems such as high dimensional integration can be viewed as the evaluation of a sum over a very large number $N$ of terms. The basic idea of the Monte Carlo method is to estimate the desired sum from a limited number $N_s$ of samples drawn from configurations $\alpha$ generated with probability $q_{\alpha}$:
\begin{equation}
M = \sum_{\alpha=1}^{N}  m_{\alpha} \approx \frac{1}{N_{S}}  \sum_{\alpha=1}^{N_S}  \frac{m_{\alpha}}{q_{\alpha}}
\label{eq:eq2}
\end{equation} 
\noindent The distribution $\{q\}$ can be uniform or could be cleverly chosen to minimize the standard deviation of the estimate \cite{bishop_pattern_2013}. In any case the standard deviation goes down as $1/\sqrt{N_s}$ and all such applications could benefit from a \textit{p-computer} to accelerate the collection of samples.

\subsection{Bayesian Network}

\label{subsec: Bayes}

A little more complicated application of a \textit{p-computer} is to problems where random numbers are generated not according to a fixed distribution, but by a distribution determined by the outputs from a previous set of $RNG's$. Consider for example the question of genetic relatedness in a family tree \cite{faria_hardware_2021, faria_implementing_2018} with each layer representing one generation. Each generation in the network in Fig. \ref{fig: fig2} (a) with $N$ nodes can be mapped to a \textit{N-bit RNG}-block feeding into a $Kernel$ which stores the conditional probability table (CPT) relating it to the next generation. The correlation between different nodes in the network can be directly measured and an average over the samples computed to yield the correct genetic correlation as shown in Fig. \ref{fig: fig2} (b). Nodes separated by $p$ generations have a correlation of $1/2^p$. The measured absolute correlation between strangers goes down to zero as $1/\sqrt{N_s}$.

This is characteristic of Monte Carlo algorithms, namely, to obtain results with accuracy $\varepsilon$ we need $N_s = 1/\varepsilon^2$ samples. The \textit{p-computer} allows us to collect samples at the rate of $N_p f_c$ \textit{= 125 MSamples per second} if $N_p=1$ and $f_c = 125$ MHz. This is about two orders of magnitude faster than what we get running the same algorithm on a Intel Xeon CPU.

How does it compare to $deterministic$ algorithms run on $CPU$? As Feynman noted in his seminal paper \cite{feynman_simulating_1982}, deterministic algorithms for problems of this type are very inefficient compared to probabilistic ones because of the need to integrate over all the unobserved nodes $\{x_B\}$ in order to calculate a property related to nodes $\{x_A\}$
\begin{equation}
P_A (x_A) = \int dx_B P(x_A,x_B)
\label{eqn: eqn3}
\end{equation}
\noindent By contrast, a \textit{p-computer} can ignore all the irrelevant nodes $\{x_b\}$ and simply look at the relevant nodes $\{x_A\}$. We used the example of genetic correlations because it is easy to relate to. But it is representative of a wide class of everyday problems involving nodes with one-way causal relationships extending from `parent' nodes to `child' nodes \cite{koller_probabilistic_2009,behin-aein_building_2016-2,yang_all-spin_2020} , all of which could benefit from a \textit{p-computer}.

\begin{figure*}
    \setlength\abovecaptionskip{-0.5\baselineskip}
    \centering
    \includegraphics[width=0.9\linewidth]{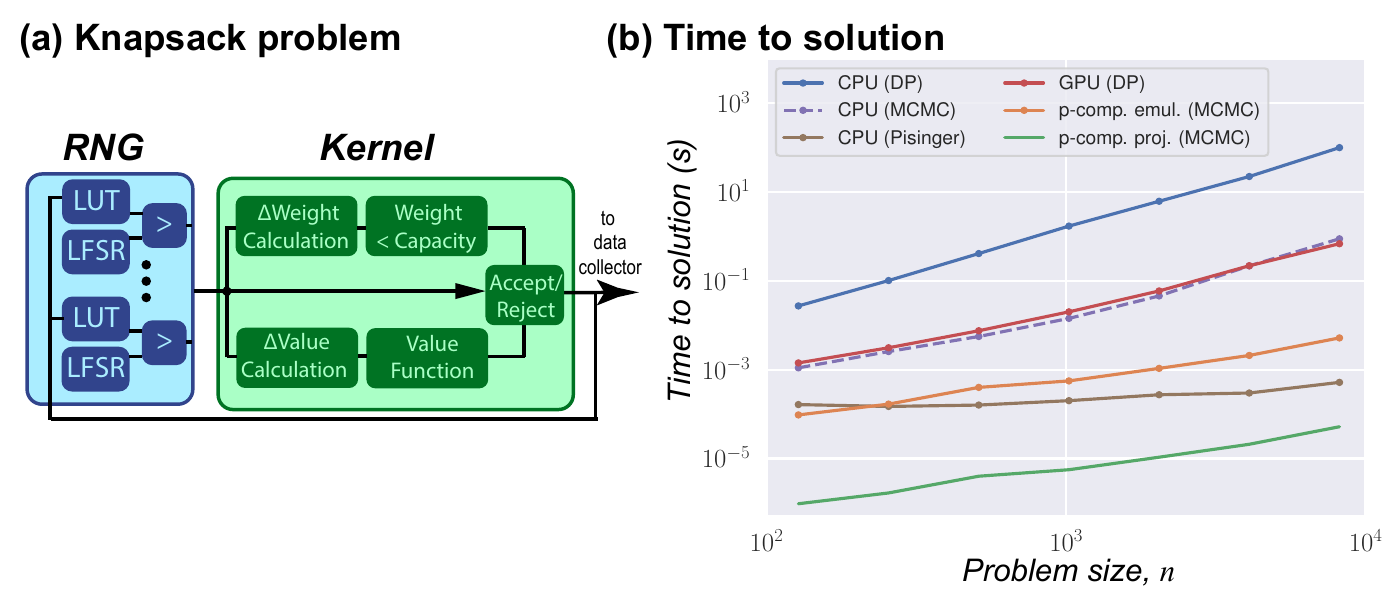} 
    \vspace{0.2in}
    \caption{\textbf{Example of MCMC - Knapsack problem: (a)} Mapping to the general \textit{p-computer} framework. \textbf{(b)} Performance of the \textit{p-computer} compared to CPU implementation of the same probabilistic algorithm and along with two well-known deterministic approaches. A deterministic algorithm like Pisinger's\cite{martello_dynamic_1999} which is optimized specifically for the Knapsack problem can outperform MCMC. But for a given MCMC algorithm, the \textit{p-computer} provides orders of magnitude improvement over the standard CPU implementation.}
    \label{fig: fig3}
\end{figure*}

\subsection{Knapsack Problem}
\label{subsec: Knapsack}

Let us now look at a problem which requires random numbers to be generated with a probability determined by the outcome from the last sample generated by the same RNG. Every RNG then requires feedback from the very Kernel that processes its output. This belongs to the broad class of problems that are labeled as \textit{Markov Chain Monte Carlo (MCMC)}. For an excellent summary and evaluation of MCMC sampling techniques we refer the reader to Ref. \cite{zhang_statistical_2021}.

The knapsack is a textbook optimization problem described in terms of a set of items, $m=1,\cdot \cdot N$, the $m^{th}$, each containing a value $v_{m}$ and weighing $w_{m}$. The problem is to figure out which items to take ($s_m=1$) and which to leave behind ($s_m=0$) such that the total value $V = \sum_{m} v_m s_m$ is a maximum, while keeping the total weight $W = \sum_{m} w_m s_m$ below a capacity $C$. We could straightforwardly map it to the $p\textendash computer$ architecture (Fig. \ref{fig: fig1}), using the $RNG$ to propose solutions $\{ s \}$ at random, and the Kernel to evaluate $V,W$ and decide to accept or reject. But this approach would take us toward the solution far too slowly. It is better to propose solutions intelligently looking at the previous accepted proposal, and making only a small change to it. For our examples we proposed a change of only two items each time.

This intelligent proposal, however, requires feedback from the kernel which can take multiple clock cycles. One could wait between proposals, but the solution is faster if instead we continue to make proposals every clock cycle in the spirit of what is referred to as \textit{multiple-try Metropolis} \cite{liu_multiple-try_2000}. The results are shown in Fig. \ref{fig: fig3}\cite{kaiser_benchmarking_2021} and compared with CPU (Intel Xeon @ 2.3GHz) and GPU (Tesla T4 @ 1.59GHz) implementations, using the probabilistic algorithm. Also shown are two efficient deterministic algorithms, one based on dynamic programming (DP), and one due to Pisinger et al. \cite{martello_dynamic_1999,kellerer_knapsack_2004}.

Note that the probabilistic algorithm (MCMC) gives solutions that are within 1\% of the correct solution, while the deterministic algorithms give the correct solution. For the Knapsack problem getting a solution that is 99\% accurate should be sufficient for most real world applications. The \textit{p-computer} provides orders of magnitude improvement over CPU implementation of the same MCMC algorithm. It is outperformed by the algorithm developed by Pisinger et al.\cite{martello_dynamic_1999,kellerer_knapsack_2004}, which is specifically optimized for the Knapsack problem. However, we note that the \textit{p-computer} projection in Fig. \ref{fig: fig3} (b) is based on utilizing better hardware like \textit{s-MTJ's} but there is also significant room for improvement of the \textit{p-computer} by optimizing the Metropolis algorithm used here and/or by adding parallel tempering \cite{swendsen_replica_1986-1,earl_parallel_2005}.

\begin{figure*}
    \setlength\abovecaptionskip{-0.5\baselineskip}
    \centering
    \includegraphics[width=0.8\linewidth]{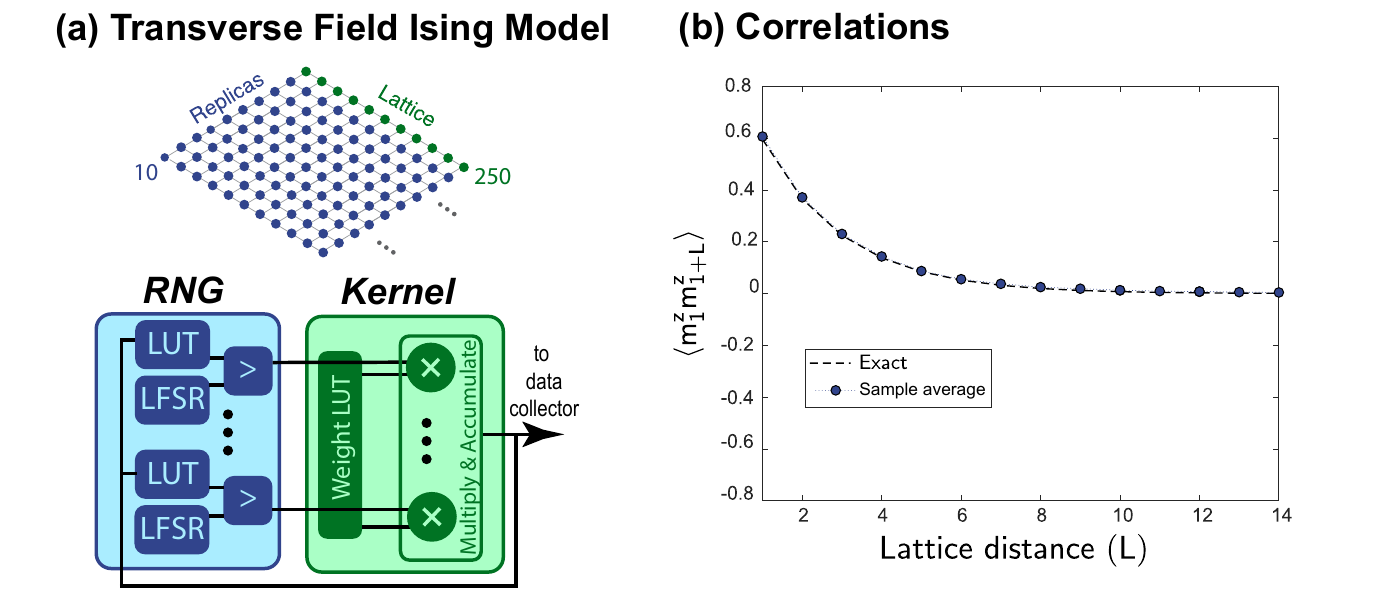} 
    \vspace{0.2in}
    \caption{\textbf{Example of Quantum Monte Carlo (QMC) - Transverse Field Ising model: (a)} Mapping to the general \textit{p-computer} framework. \textbf{(b)} Solving the transverse Ising model for quantum annealing. Subfigure (b) is adapted from B. Sutton, R. Faria, L. A. Ghantasala, R. Jaiswal, K. Y. Camsari and S. Datta, IEEE Access, vol. 8, pp. 157238-157252, 2020 licensed under a Creative Commons Attribution \href{https://creativecommons.org/licenses/by/4.0/}{(CC BY)} license\cite{sutton_autonomous_2020}.}
    \label{fig: fig4}
\end{figure*}
\subsection{Ising model}

Another widely used model for optimization within MCMC is based on the concept of \textit{Boltzmann machines (BM)} defined by an energy function $E$ from which one can calculate the synaptic function $I_i$
\begin{equation}
I_i= \beta (E(s_i=0) - E(s_i=1)),
\label{eqn:Synapse}
\end{equation}
\noindent that can be used to guide the sample generation from each \textit{RNG `i'} in sequence\cite{geman_stochastic_1984-2} according to Eq. (\ref{eqn: eqn1}). Alternatively the sample generation from each RNG can be fixed and the synaptic function used to decide whether to accept or reject it within a Metropolis-Hastings framework \cite{hastings_monte_1970}. Either way, samples will be generated with probabilities $P_{\alpha} \sim exp(-\beta E_{\alpha})$. We can solve optimization problems by identifying $E$ with the negative of the cost function that we are seeking to minimize. Using a large $\beta$ we can ensure that the probability is nearly $1$ for the configuration with the minimum value of E. 

In principle, the energy function is arbitrary, but much of the work is based on quadratic energy functions defined by a connection matrix $W_{ij}$ and a bias vector $h_i$ (see for example \citep{goto_combinatorial_2019,aramon_physics-inspired_2019,yamamoto_73_2020,yamaoka_243_2015,ahmed_probabilistic_2020,sutton_autonomous_2020,patel_ising_2020,cai_power-efficient_2020,dutta_ising_2021}):
\begin{equation}
E=-\sum_{ij} W_{ij} s_i s_j -\sum_i h_i s_i,
\label{eqn:Ising}
\end{equation}
For this quadratic energy function, Eq. (\ref{eqn:Synapse}) gives $I_i = \beta \big(\sum_{j} W_{ij} s_j + h_i \big)$, so that the $Kernel$ has to perform a multiply and accumulate operation as shown in Fig. \ref{fig: fig4} (a). We refer the reader to Sutton et al. \cite{sutton_autonomous_2020} for an example of the \textit{max-cut} optimization problem on a two-dimensional $90 \times 90$ array implemented with a \textit{p-computer}.

Eq. (\ref{eqn:Synapse}), however, is more generally applicable even if the energy expression is more complicated, or given by a table. The $Kernel$ can be modified accordingly. For an example of a energy function with fourth order terms implemented on an eight bit \textit{p-computer}, we refer the reader to Borders et al.\cite{borders_integer_2019}.

A wide variety of problems can be mapped onto the BM with an appropriate choice of the energy function. For example, we could generate samples from a desired probability distribution $P$, by choosing $\beta E = - \ell n P$. Another example is the implementation of logic gates by defining $E$ to be zero for all $\{ s \}$ that belong to the truth table, and have some positive value for those that do not \cite{camsari_stochastic_2017}. Unlike standard digital logic, such a BM-based implementation would provide \textit{invertible logic} that not only provides the output for a given input, but also generates all possible inputs corresponding to a specified output\cite{camsari_stochastic_2017,lv_experimental_2019,aadit_computing_2021}.

\subsection{Quantum Monte Carlo}
Finally let us briefly describe the feasibility of using \textit{p-computers} to emulate quantum or q-computers. A q-computer is based on $qubits$ that are neither $0$ or $1$, but are described by a complex wavefunction whose squared magnitude gives the probability of measuring either a $0$ or a $1$. The state of an $n \textendash qubit$ computer is described by a wavefunction $\{\psi\}$ with $2^{n}$ complex components, one for $\textit{each possible configuration}$ of the $n$ qubits.

In gate-based quantum computing (GQC) a set of qubits is placed in a known state at time $t$, operated on with $d$ quantum gates to manipulate the wavefunction through unitary transformations $[U^{(i)}]$
\begin{equation}
\{\psi(t+d)\} = [U^{(d)}] \cdot \cdot \ \ \cdot \cdot [U^{(1)}] \{\psi(t)\} \ \ \ \ \ \ \ \ (GQC) 
     \label{eq1a}
\end{equation}
\noindent and measurements are made to obtain results with probabilities given by the squared magnitudes of the final wavefunctions. From the rules of matrix multiplication, the final wavefunction can be written as a sum over a very large number of terms: 
\begin{equation}
\psi_m(t+d) = \sum_{i, \cdot \cdot j,k} U_{m,i}^{(d)} \cdot \cdot \ \ \cdot \cdot \ U_{j,k}^{(1)} \  \psi_{k}(t) 
     \label{eq1b}
\end{equation}
\noindent Conceptually we could represent a system of $n$ qubits and $d$ gates with a system of $(n \times d)$ \textit{p-bits} with $2^{nd}$ states which label the $2^{nd}$  terms in the summation in Eq.(\ref{eq1b}) \cite{chowdhury_emulating_2020}. Each of these terms is often referred to as a \textit{Feynman path} and what we want is the sum of the amplitudes of all such paths:
\begin{equation}
\psi_m(t+d) =  \sum_{\alpha=1}^{2^{nd}}  A_m^{(\alpha)} 
\label{eq:eq10}
\end{equation} 

\noindent The essential idea of \textit{quantum Monte Carlo} is to estimate this enormous sum from a few suitably chosen samples, not unlike the simple Monte Carlo stated earlier in Eq. (\ref{eq:eq2}). What makes it more difficult, however, is the so-called \textit{sign problem} \cite{troyer_computational_2005} which can be understood intuitively as follows. If all the quantities $A_m^{(\alpha)}$ are positive then it is relatively easy to estimate the sum from a few samples. But if some are positive while some are negative with lots of cancellations, then many more samples will be required. The same is true if the quantities are complex quantities that cancel each other.

The matrices $U$ that appear in GQC are unitary with complex elements which often leads to significant cancellation of Feynman paths, except in special cases when there may be complete constructive interference. In general this could make it necessary to use large numbers of samples for accurate estimation. A noiseless quantum computer would not have this problem, since qubits intuitively perform the entire sum exactly and yield samples according to the squared magnitude of the resulting wavefunction. However, real world quantum computers have noise and \textit{p-computers} could be competitive for many problems.

Adiabatic quantum computing (AQC) operates on very different physical principles but its mathematical description can also be viewed as summing the Feynman paths representing the multiplication of $r$ matrices:
\begin{equation}
[e^{-\beta H/r}] \cdot \cdot \ \ \cdot \cdot [e^{-\beta H/r}]   \ \ \ \ \ \ \ \ (AQC) 
     \label{eq1c}
\end{equation}
\noindent This is based on the Suzuki-Trotter method described in Camsari et al. \cite{camsari_scalable_2019}, where the number of replicas, $r$, is chosen large enough to ensure that if $H=H_1+H_2$, one can approximately write $e^{ -\beta H/r} \approx e^{-\beta H_1/r} e^{-\beta H_2/r}$. The matrices $e^{ -\beta H/r}$ in AQC are Hermitian, and  their elements can be all positive\cite{bravyi_complexity_2008}. A special class of Hamiltonians $H$ having this property is called \textit{stoquastic} and there is no sign problem since the amplitudes $A_m^{(\alpha)}$ in the Feynman sum in Eq. (\ref{eq:eq10}) all have the same sign.

An example of such a stoquastic Hamiltonian is the transverse field Ising model (TFIM) commonly used for \textit{quantum annealing} where a transverse field which is quantum in nature is introduced and slowly reduced to zero to recover the original classical problem. Fig. \ref{fig: fig4} adapted from Sutton et al.\cite{sutton_autonomous_2020}, shows a \textit{n = 250 qubit} problem mapped to a 2-D lattice of $250 \times 10 = 2500$ \textit{p-bits} using $r=10$ replicas to calculated average correlations between the z-directed spins on lattice sites separated by $L$. Very accurate results are obtained using $N_s$ = $10^5$ samples. However,these samples were suitably spaced to ensure their independence, which is an important concern in problems involving feedback.

Finally we note that quantum Monte Carlo methods, both GQC and AQC, involve selective summing of Feynman paths to evaluate matrix products. As such we might expect conceptual overlap with the very active field of randomized algorithms for linear algebra \cite{drineas_fast_2006-1,drineas_fast_2006-2}, though the two fields seem very distinct at this time.

\section{Concluding Remarks}
In summary, we have presented a generic architecture for a \textit{p-computer} based on \textit{p-bits} which take on values $0$ and $1$ with controlled probabilities, and can be implemented with specialized compact energy-efficient hardware. We emulate systems with thousands of \textit{p-bits} to show that they can significantly accelerate the implementation of randomized algorithms that are widely used for many applications\cite{buluc_randomized_2021}. A few prototypical examples are presented such as Bayesian networks, optimization, Ising models and quantum Monte Carlo. 

\begin{acknowledgments}
The authors are grateful to Behtash Behin-Aein for helpful discussions and advice. We also thank Kerem Camsari and Shuvro Chowdhury for their feedback on the manuscript. The contents are based on the work done over the last 5-10 years in our group, some of which has been cited here, and it is a pleasure to acknowledge all who have contributed to our understanding. This work was supported in part by ASCENT, one of six centers in JUMP, a Semiconductor Research Corporation (SRC) program sponsored by DARPA.
\end{acknowledgments}

\section*{Data Availability}
\vspace{-0.1in}
The data that support the findings of this study are available from the corresponding author upon reasonable request.

\vspace{-0.1in}

\section*{Author Declarations}
\subsection*{Conflict of interests}
\vspace{-0.1in}
One of the authors (SD) has a financial interest in Ludwig Computing.

\bibliography{library}

\begin{thebibliography}{53}%
\makeatletter
\providecommand \@ifxundefined [1]{%
 \@ifx{#1\undefined}
}%
\providecommand \@ifnum [1]{%
 \ifnum #1\expandafter \@firstoftwo
 \else \expandafter \@secondoftwo
 \fi
}%
\providecommand \@ifx [1]{%
 \ifx #1\expandafter \@firstoftwo
 \else \expandafter \@secondoftwo
 \fi
}%
\providecommand \natexlab [1]{#1}%
\providecommand \enquote  [1]{``#1''}%
\providecommand \bibnamefont  [1]{#1}%
\providecommand \bibfnamefont [1]{#1}%
\providecommand \citenamefont [1]{#1}%
\providecommand \href@noop [0]{\@secondoftwo}%
\providecommand \href [0]{\begingroup \@sanitize@url \@href}%
\providecommand \@href[1]{\@@startlink{#1}\@@href}%
\providecommand \@@href[1]{\endgroup#1\@@endlink}%
\providecommand \@sanitize@url [0]{\catcode `\\12\catcode `\$12\catcode
  `\&12\catcode `\#12\catcode `\^12\catcode `\_12\catcode `\%12\relax}%
\providecommand \@@startlink[1]{}%
\providecommand \@@endlink[0]{}%
\providecommand \url  [0]{\begingroup\@sanitize@url \@url }%
\providecommand \@url [1]{\endgroup\@href {#1}{\urlprefix }}%
\providecommand \urlprefix  [0]{URL }%
\providecommand \Eprint [0]{\href }%
\providecommand \doibase [0]{http://dx.doi.org/}%
\providecommand \selectlanguage [0]{\@gobble}%
\providecommand \bibinfo  [0]{\@secondoftwo}%
\providecommand \bibfield  [0]{\@secondoftwo}%
\providecommand \translation [1]{[#1]}%
\providecommand \BibitemOpen [0]{}%
\providecommand \bibitemStop [0]{}%
\providecommand \bibitemNoStop [0]{.\EOS\space}%
\providecommand \EOS [0]{\spacefactor3000\relax}%
\providecommand \BibitemShut  [1]{\csname bibitem#1\endcsname}%
\let\auto@bib@innerbib\@empty
\bibitem [{\citenamefont {Feynman}(1982)}]{feynman_simulating_1982}%
  \BibitemOpen
  \bibfield  {author} {\bibinfo {author} {\bibfnamefont {R.~P.}\ \bibnamefont
  {Feynman}},\ }\bibfield  {title} {\enquote {\bibinfo {title} {Simulating
  physics with computers},}\ }\href@noop {} {\bibfield  {journal} {\bibinfo
  {journal} {Int. J. Theor. Phys}\ }\textbf {\bibinfo {volume} {21}} (\bibinfo
  {year} {1982})}\BibitemShut {NoStop}%
\bibitem [{\citenamefont {Cipra}(2000)}]{cipra_best_2000}%
  \BibitemOpen
  \bibfield  {author} {\bibinfo {author} {\bibfnamefont {B.~A.}\ \bibnamefont
  {Cipra}},\ }\bibfield  {title} {\enquote {\bibinfo {title} {The {{Best}} of
  the 20th {{Century}}: Editors {{Name Top}} 10 {{Algorithms}}},}\ }\href@noop
  {} {\bibfield  {journal} {\bibinfo  {journal} {SIAM News}\ }\textbf {\bibinfo
  {volume} {33}},\ \bibinfo {pages} {1--2} (\bibinfo {year}
  {2000})}\BibitemShut {NoStop}%
\bibitem [{\citenamefont {Camsari}, \citenamefont {Salahuddin},\ and\
  \citenamefont {Datta}(2017)}]{camsari_implementing_2017}%
  \BibitemOpen
  \bibfield  {author} {\bibinfo {author} {\bibfnamefont {K.~Y.}\ \bibnamefont
  {Camsari}}, \bibinfo {author} {\bibfnamefont {S.}~\bibnamefont {Salahuddin}},
  \ and\ \bibinfo {author} {\bibfnamefont {S.}~\bibnamefont {Datta}},\
  }\bibfield  {title} {\enquote {\bibinfo {title} {Implementing p-bits with
  {{Embedded MTJ}}},}\ }\href@noop {} {\bibfield  {journal} {\bibinfo
  {journal} {IEEE Electron Device Letters}\ }\textbf {\bibinfo {volume} {38}},\
  \bibinfo {pages} {1767--1770} (\bibinfo {year} {2017})}\BibitemShut {NoStop}%
\bibitem [{\citenamefont {Kaiser}\ \emph
  {et~al.}(2021{\natexlab{a}})\citenamefont {Kaiser}, \citenamefont {Jaiswal},
  \citenamefont {{Behin-Aein}},\ and\ \citenamefont
  {Datta}}]{kaiser_benchmarking_2021}%
  \BibitemOpen
  \bibfield  {author} {\bibinfo {author} {\bibfnamefont {J.}~\bibnamefont
  {Kaiser}}, \bibinfo {author} {\bibfnamefont {R.}~\bibnamefont {Jaiswal}},
  \bibinfo {author} {\bibfnamefont {B.}~\bibnamefont {{Behin-Aein}}}, \ and\
  \bibinfo {author} {\bibfnamefont {S.}~\bibnamefont {Datta}},\ }\bibfield
  {title} {\enquote {\bibinfo {title} {Benchmarking a {{Probabilistic
  Coprocessor}}},}\ }\href@noop {} {\bibfield  {journal} {\bibinfo  {journal}
  {arXiv:2109.14801 [cond-mat]}\ } (\bibinfo {year} {2021}{\natexlab{a}})},\
  \Eprint {http://arxiv.org/abs/2109.14801} {arXiv:2109.14801 [cond-mat]}
  \BibitemShut {NoStop}%
\bibitem [{\citenamefont {Swendsen}\ and\ \citenamefont
  {Wang}(1986)}]{swendsen_replica_1986-1}%
  \BibitemOpen
  \bibfield  {author} {\bibinfo {author} {\bibfnamefont {R.}~\bibnamefont
  {Swendsen}}\ and\ \bibinfo {author} {\bibfnamefont {J.-S.}\ \bibnamefont
  {Wang}},\ }\bibfield  {title} {\enquote {\bibinfo {title} {Replica {{Monte
  Carlo Simulation}} of {{Spin}}-{{Glasses}}},}\ }\href {\doibase
  10.1103/PhysRevLett.57.2607} {\bibfield  {journal} {\bibinfo  {journal}
  {Physical review letters}\ }\textbf {\bibinfo {volume} {57}},\ \bibinfo
  {pages} {2607--2609} (\bibinfo {year} {1986})}\BibitemShut {NoStop}%
\bibitem [{\citenamefont {Earl}\ and\ \citenamefont
  {Deem}(2005)}]{earl_parallel_2005}%
  \BibitemOpen
  \bibfield  {author} {\bibinfo {author} {\bibfnamefont {D.~J.}\ \bibnamefont
  {Earl}}\ and\ \bibinfo {author} {\bibfnamefont {M.~W.}\ \bibnamefont
  {Deem}},\ }\bibfield  {title} {\enquote {\bibinfo {title} {Parallel
  tempering: Theory, applications, and new perspectives},}\ }\href {\doibase
  10.1039/B509983H} {\bibfield  {journal} {\bibinfo  {journal} {Physical
  Chemistry Chemical Physics}\ }\textbf {\bibinfo {volume} {7}},\ \bibinfo
  {pages} {3910--3916} (\bibinfo {year} {2005})}\BibitemShut {NoStop}%
\bibitem [{\citenamefont {Ko}\ \emph {et~al.}(2019)\citenamefont {Ko},
  \citenamefont {Chai}, \citenamefont {Rutenbar}, \citenamefont {Brooks},\ and\
  \citenamefont {Wei}}]{ko_accelerating_2019}%
  \BibitemOpen
  \bibfield  {author} {\bibinfo {author} {\bibfnamefont {G.~G.}\ \bibnamefont
  {Ko}}, \bibinfo {author} {\bibfnamefont {Y.}~\bibnamefont {Chai}}, \bibinfo
  {author} {\bibfnamefont {R.~A.}\ \bibnamefont {Rutenbar}}, \bibinfo {author}
  {\bibfnamefont {D.}~\bibnamefont {Brooks}}, \ and\ \bibinfo {author}
  {\bibfnamefont {G.-Y.}\ \bibnamefont {Wei}},\ }\bibfield  {title} {\enquote
  {\bibinfo {title} {Accelerating {{Bayesian Inference}} on {{Structured Graphs
  Using Parallel Gibbs Sampling}}},}\ }in\ \href {\doibase
  10.1109/FPL.2019.00033} {\emph {\bibinfo {booktitle} {2019 29th
  {{International Conference}} on {{Field Programmable Logic}} and
  {{Applications}} ({{FPL}})}}}\ (\bibinfo  {publisher} {{IEEE}},\ \bibinfo
  {address} {{Barcelona, Spain}},\ \bibinfo {year} {2019})\ pp.\ \bibinfo
  {pages} {159--165}\BibitemShut {NoStop}%
\bibitem [{\citenamefont {Sutton}\ \emph {et~al.}(2020)\citenamefont {Sutton},
  \citenamefont {Faria}, \citenamefont {Ghantasala}, \citenamefont {Jaiswal},
  \citenamefont {Camsari},\ and\ \citenamefont
  {Datta}}]{sutton_autonomous_2020}%
  \BibitemOpen
  \bibfield  {author} {\bibinfo {author} {\bibfnamefont {B.}~\bibnamefont
  {Sutton}}, \bibinfo {author} {\bibfnamefont {R.}~\bibnamefont {Faria}},
  \bibinfo {author} {\bibfnamefont {L.~A.}\ \bibnamefont {Ghantasala}},
  \bibinfo {author} {\bibfnamefont {R.}~\bibnamefont {Jaiswal}}, \bibinfo
  {author} {\bibfnamefont {K.~Y.}\ \bibnamefont {Camsari}}, \ and\ \bibinfo
  {author} {\bibfnamefont {S.}~\bibnamefont {Datta}},\ }\bibfield  {title}
  {\enquote {\bibinfo {title} {Autonomous {{Probabilistic Coprocessing}} with
  {{Petaflips}} per {{Second}}},}\ }\href {\doibase
  10.1109/ACCESS.2020.3018682} {\bibfield  {journal} {\bibinfo  {journal} {IEEE
  Access}\ ,\ \bibinfo {pages} {1--1}} (\bibinfo {year} {2020})}\BibitemShut
  {NoStop}%
\bibitem [{Note1()}]{Note1}%
  \BibitemOpen
  \bibinfo {note} {Here, we assume that every RNG-Kernel unit gives one sample
  per clock cycle. There could be cases where multiple samples could be
  extracted from one unit per clock cycle.}\BibitemShut {Stop}%
\bibitem [{\citenamefont {Goto}, \citenamefont {Tatsumura},\ and\ \citenamefont
  {Dixon}(2019)}]{goto_combinatorial_2019}%
  \BibitemOpen
  \bibfield  {author} {\bibinfo {author} {\bibfnamefont {H.}~\bibnamefont
  {Goto}}, \bibinfo {author} {\bibfnamefont {K.}~\bibnamefont {Tatsumura}}, \
  and\ \bibinfo {author} {\bibfnamefont {A.~R.}\ \bibnamefont {Dixon}},\
  }\bibfield  {title} {\enquote {\bibinfo {title} {Combinatorial optimization
  by simulating adiabatic bifurcations in nonlinear {{Hamiltonian}} systems},}\
  }\href {\doibase 10.1126/sciadv.aav2372} {\bibfield  {journal} {\bibinfo
  {journal} {Science Advances}\ }\textbf {\bibinfo {volume} {5}},\ \bibinfo
  {pages} {eaav2372} (\bibinfo {year} {2019})}\BibitemShut {NoStop}%
\bibitem [{\citenamefont {Aramon}\ \emph {et~al.}(2019)\citenamefont {Aramon},
  \citenamefont {Rosenberg}, \citenamefont {Valiante}, \citenamefont
  {Miyazawa}, \citenamefont {Tamura},\ and\ \citenamefont
  {Katzgraber}}]{aramon_physics-inspired_2019}%
  \BibitemOpen
  \bibfield  {author} {\bibinfo {author} {\bibfnamefont {M.}~\bibnamefont
  {Aramon}}, \bibinfo {author} {\bibfnamefont {G.}~\bibnamefont {Rosenberg}},
  \bibinfo {author} {\bibfnamefont {E.}~\bibnamefont {Valiante}}, \bibinfo
  {author} {\bibfnamefont {T.}~\bibnamefont {Miyazawa}}, \bibinfo {author}
  {\bibfnamefont {H.}~\bibnamefont {Tamura}}, \ and\ \bibinfo {author}
  {\bibfnamefont {H.~G.}\ \bibnamefont {Katzgraber}},\ }\bibfield  {title}
  {\enquote {\bibinfo {title} {Physics-{{Inspired Optimization}} for
  {{Quadratic Unconstrained Problems Using}} a {{Digital Annealer}}},}\ }\href
  {\doibase 10.3389/fphy.2019.00048} {\bibfield  {journal} {\bibinfo  {journal}
  {Frontiers in Physics}\ }\textbf {\bibinfo {volume} {7}} (\bibinfo {year}
  {2019}),\ 10.3389/fphy.2019.00048}\BibitemShut {NoStop}%
\bibitem [{\citenamefont {Yamamoto}\ \emph {et~al.}(2020)\citenamefont
  {Yamamoto}, \citenamefont {Ando}, \citenamefont {Mertig}, \citenamefont
  {Takemoto}, \citenamefont {Yamaoka}, \citenamefont {Teramoto}, \citenamefont
  {Sakai}, \citenamefont {{Takamaeda-Yamazaki}},\ and\ \citenamefont
  {Motomura}}]{yamamoto_73_2020}%
  \BibitemOpen
  \bibfield  {author} {\bibinfo {author} {\bibfnamefont {K.}~\bibnamefont
  {Yamamoto}}, \bibinfo {author} {\bibfnamefont {K.}~\bibnamefont {Ando}},
  \bibinfo {author} {\bibfnamefont {N.}~\bibnamefont {Mertig}}, \bibinfo
  {author} {\bibfnamefont {T.}~\bibnamefont {Takemoto}}, \bibinfo {author}
  {\bibfnamefont {M.}~\bibnamefont {Yamaoka}}, \bibinfo {author} {\bibfnamefont
  {H.}~\bibnamefont {Teramoto}}, \bibinfo {author} {\bibfnamefont
  {A.}~\bibnamefont {Sakai}}, \bibinfo {author} {\bibfnamefont
  {S.}~\bibnamefont {{Takamaeda-Yamazaki}}}, \ and\ \bibinfo {author}
  {\bibfnamefont {M.}~\bibnamefont {Motomura}},\ }\bibfield  {title} {\enquote
  {\bibinfo {title} {7.3 {{STATICA}}: A 512-{{Spin}} 0.{{25M}}-{{Weight
  Full}}-{{Digital Annealing Processor}} with a {{Near}}-{{Memory
  All}}-{{Spin}}-{{Updates}}-at-{{Once Architecture}} for {{Combinatorial
  Optimization}} with {{Complete Spin}}-{{Spin Interactions}}},}\ }in\ \href
  {\doibase 10.1109/ISSCC19947.2020.9062965} {\emph {\bibinfo {booktitle} {2020
  {{IEEE International Solid}}- {{State Circuits Conference}} - ({{ISSCC}})}}}\
  (\bibinfo  {publisher} {{IEEE}},\ \bibinfo {address} {{San Francisco, CA,
  USA}},\ \bibinfo {year} {2020})\ pp.\ \bibinfo {pages} {138--140}\BibitemShut
  {NoStop}%
\bibitem [{\citenamefont {Yamaoka}\ \emph {et~al.}(2015)\citenamefont
  {Yamaoka}, \citenamefont {Yoshimura}, \citenamefont {Hayashi}, \citenamefont
  {Okuyama}, \citenamefont {Aoki},\ and\ \citenamefont
  {Mizuno}}]{yamaoka_243_2015}%
  \BibitemOpen
  \bibfield  {author} {\bibinfo {author} {\bibfnamefont {M.}~\bibnamefont
  {Yamaoka}}, \bibinfo {author} {\bibfnamefont {C.}~\bibnamefont {Yoshimura}},
  \bibinfo {author} {\bibfnamefont {M.}~\bibnamefont {Hayashi}}, \bibinfo
  {author} {\bibfnamefont {T.}~\bibnamefont {Okuyama}}, \bibinfo {author}
  {\bibfnamefont {H.}~\bibnamefont {Aoki}}, \ and\ \bibinfo {author}
  {\bibfnamefont {H.}~\bibnamefont {Mizuno}},\ }\bibfield  {title} {\enquote
  {\bibinfo {title} {24.3 20k-spin {{Ising}} chip for combinational
  optimization problem with {{CMOS}} annealing},}\ }in\ \href {\doibase
  10.1109/ISSCC.2015.7063111} {\emph {\bibinfo {booktitle} {2015 {{IEEE
  International Solid}}-{{State Circuits Conference}} - ({{ISSCC}}) {{Digest}}
  of {{Technical Papers}}}}}\ (\bibinfo {year} {2015})\ pp.\ \bibinfo {pages}
  {1--3}\BibitemShut {NoStop}%
\bibitem [{\citenamefont {Ahmed}, \citenamefont {Chiu},\ and\ \citenamefont
  {Kim}(2020)}]{ahmed_probabilistic_2020}%
  \BibitemOpen
  \bibfield  {author} {\bibinfo {author} {\bibfnamefont {I.}~\bibnamefont
  {Ahmed}}, \bibinfo {author} {\bibfnamefont {P.-W.}\ \bibnamefont {Chiu}}, \
  and\ \bibinfo {author} {\bibfnamefont {C.~H.}\ \bibnamefont {Kim}},\
  }\bibfield  {title} {\enquote {\bibinfo {title} {A {{Probabilistic
  Self}}-{{Annealing Compute Fabric Based}} on 560 {{Hexagonally Coupled Ring
  Oscillators}} for {{Solving Combinatorial Optimization Problems}}},}\ }in\
  \href {\doibase 10.1109/VLSICircuits18222.2020.9162869} {\emph {\bibinfo
  {booktitle} {2020 {{IEEE Symposium}} on {{VLSI Circuits}}}}}\ (\bibinfo
  {publisher} {{IEEE}},\ \bibinfo {address} {{Honolulu, HI, USA}},\ \bibinfo
  {year} {2020})\ pp.\ \bibinfo {pages} {1--2}\BibitemShut {NoStop}%
\bibitem [{\citenamefont {Patel}\ \emph {et~al.}(2020)\citenamefont {Patel},
  \citenamefont {Chen}, \citenamefont {Canoza},\ and\ \citenamefont
  {Salahuddin}}]{patel_ising_2020}%
  \BibitemOpen
  \bibfield  {author} {\bibinfo {author} {\bibfnamefont {S.}~\bibnamefont
  {Patel}}, \bibinfo {author} {\bibfnamefont {L.}~\bibnamefont {Chen}},
  \bibinfo {author} {\bibfnamefont {P.}~\bibnamefont {Canoza}}, \ and\ \bibinfo
  {author} {\bibfnamefont {S.}~\bibnamefont {Salahuddin}},\ }\bibfield  {title}
  {\enquote {\bibinfo {title} {Ising {{Model Optimization Problems}} on a
  {{FPGA Accelerated Restricted Boltzmann Machine}}},}\ }\href@noop {}
  {\bibfield  {journal} {\bibinfo  {journal} {arXiv:2008.04436 [physics]}\ }
  (\bibinfo {year} {2020})},\ \Eprint {http://arxiv.org/abs/2008.04436}
  {arXiv:2008.04436 [physics]} \BibitemShut {NoStop}%
\bibitem [{\citenamefont {Cai}\ \emph {et~al.}(2020)\citenamefont {Cai},
  \citenamefont {Kumar}, \citenamefont {Van~Vaerenbergh}, \citenamefont
  {Sheng}, \citenamefont {Liu}, \citenamefont {Li}, \citenamefont {Liu},
  \citenamefont {Foltin}, \citenamefont {Yu}, \citenamefont {Xia},
  \citenamefont {Yang}, \citenamefont {Beausoleil}, \citenamefont {Lu},\ and\
  \citenamefont {Strachan}}]{cai_power-efficient_2020}%
  \BibitemOpen
  \bibfield  {author} {\bibinfo {author} {\bibfnamefont {F.}~\bibnamefont
  {Cai}}, \bibinfo {author} {\bibfnamefont {S.}~\bibnamefont {Kumar}}, \bibinfo
  {author} {\bibfnamefont {T.}~\bibnamefont {Van~Vaerenbergh}}, \bibinfo
  {author} {\bibfnamefont {X.}~\bibnamefont {Sheng}}, \bibinfo {author}
  {\bibfnamefont {R.}~\bibnamefont {Liu}}, \bibinfo {author} {\bibfnamefont
  {C.}~\bibnamefont {Li}}, \bibinfo {author} {\bibfnamefont {Z.}~\bibnamefont
  {Liu}}, \bibinfo {author} {\bibfnamefont {M.}~\bibnamefont {Foltin}},
  \bibinfo {author} {\bibfnamefont {S.}~\bibnamefont {Yu}}, \bibinfo {author}
  {\bibfnamefont {Q.}~\bibnamefont {Xia}}, \bibinfo {author} {\bibfnamefont
  {J.~J.}\ \bibnamefont {Yang}}, \bibinfo {author} {\bibfnamefont
  {R.}~\bibnamefont {Beausoleil}}, \bibinfo {author} {\bibfnamefont {W.~D.}\
  \bibnamefont {Lu}}, \ and\ \bibinfo {author} {\bibfnamefont {J.~P.}\
  \bibnamefont {Strachan}},\ }\bibfield  {title} {\enquote {\bibinfo {title}
  {Power-efficient combinatorial optimization using intrinsic noise in
  memristor {{Hopfield}} neural networks},}\ }\href {\doibase
  10.1038/s41928-020-0436-6} {\bibfield  {journal} {\bibinfo  {journal} {Nature
  Electronics}\ }\textbf {\bibinfo {volume} {3}},\ \bibinfo {pages} {409--418}
  (\bibinfo {year} {2020})}\BibitemShut {NoStop}%
\bibitem [{\citenamefont {Dutta}\ \emph {et~al.}(2021)\citenamefont {Dutta},
  \citenamefont {Khanna}, \citenamefont {Assoa}, \citenamefont {Paik},
  \citenamefont {Schlom}, \citenamefont {Toroczkai}, \citenamefont
  {Raychowdhury},\ and\ \citenamefont {Datta}}]{dutta_ising_2021}%
  \BibitemOpen
  \bibfield  {author} {\bibinfo {author} {\bibfnamefont {S.}~\bibnamefont
  {Dutta}}, \bibinfo {author} {\bibfnamefont {A.}~\bibnamefont {Khanna}},
  \bibinfo {author} {\bibfnamefont {A.~S.}\ \bibnamefont {Assoa}}, \bibinfo
  {author} {\bibfnamefont {H.}~\bibnamefont {Paik}}, \bibinfo {author}
  {\bibfnamefont {D.~G.}\ \bibnamefont {Schlom}}, \bibinfo {author}
  {\bibfnamefont {Z.}~\bibnamefont {Toroczkai}}, \bibinfo {author}
  {\bibfnamefont {A.}~\bibnamefont {Raychowdhury}}, \ and\ \bibinfo {author}
  {\bibfnamefont {S.}~\bibnamefont {Datta}},\ }\bibfield  {title} {\enquote
  {\bibinfo {title} {An {{Ising Hamiltonian}} solver based on coupled
  stochastic phase-transition nano-oscillators},}\ }\href {\doibase
  10.1038/s41928-021-00616-7} {\bibfield  {journal} {\bibinfo  {journal}
  {Nature Electronics}\ }\textbf {\bibinfo {volume} {4}},\ \bibinfo {pages}
  {502--512} (\bibinfo {year} {2021})}\BibitemShut {NoStop}%
\bibitem [{\citenamefont {Camsari}\ and\ \citenamefont
  {Datta}(2021)}]{camsari_dialogue_2021}%
  \BibitemOpen
  \bibfield  {author} {\bibinfo {author} {\bibfnamefont {K.}~\bibnamefont
  {Camsari}}\ and\ \bibinfo {author} {\bibfnamefont {S.}~\bibnamefont
  {Datta}},\ }\bibfield  {title} {\enquote {\bibinfo {title} {Dialogue
  {{Concerning}} the {{Two Chief Computing Systems}}: Imagine yourself on a
  flight talking to an engineer about a scheme that straddles classical and
  quantum},}\ }\href {\doibase 10.1109/MSPEC.2021.9393992} {\bibfield
  {journal} {\bibinfo  {journal} {IEEE Spectrum}\ }\textbf {\bibinfo {volume}
  {58}},\ \bibinfo {pages} {30--35} (\bibinfo {year} {2021})}\BibitemShut
  {NoStop}%
\bibitem [{\citenamefont {Troyer}\ and\ \citenamefont
  {Wiese}(2005)}]{troyer_computational_2005}%
  \BibitemOpen
  \bibfield  {author} {\bibinfo {author} {\bibfnamefont {M.}~\bibnamefont
  {Troyer}}\ and\ \bibinfo {author} {\bibfnamefont {U.-J.}\ \bibnamefont
  {Wiese}},\ }\bibfield  {title} {\enquote {\bibinfo {title} {Computational
  {{Complexity}} and {{Fundamental Limitations}} to {{Fermionic Quantum Monte
  Carlo Simulations}}},}\ }\href {\doibase 10.1103/PhysRevLett.94.170201}
  {\bibfield  {journal} {\bibinfo  {journal} {Physical Review Letters}\
  }\textbf {\bibinfo {volume} {94}},\ \bibinfo {pages} {170201} (\bibinfo
  {year} {2005})}\BibitemShut {NoStop}%
\bibitem [{\citenamefont {Li}\ \emph {et~al.}(2020)\citenamefont {Li},
  \citenamefont {Hou}, \citenamefont {Yan}, \citenamefont {Liu}, \citenamefont
  {Yang},\ and\ \citenamefont {Sun}}]{li_chiplet_2020}%
  \BibitemOpen
  \bibfield  {author} {\bibinfo {author} {\bibfnamefont {T.}~\bibnamefont
  {Li}}, \bibinfo {author} {\bibfnamefont {J.}~\bibnamefont {Hou}}, \bibinfo
  {author} {\bibfnamefont {J.}~\bibnamefont {Yan}}, \bibinfo {author}
  {\bibfnamefont {R.}~\bibnamefont {Liu}}, \bibinfo {author} {\bibfnamefont
  {H.}~\bibnamefont {Yang}}, \ and\ \bibinfo {author} {\bibfnamefont
  {Z.}~\bibnamefont {Sun}},\ }\bibfield  {title} {\enquote {\bibinfo {title}
  {Chiplet {{Heterogeneous Integration Technology}}\textemdash{{Status}} and
  {{Challenges}}},}\ }\href {\doibase 10.3390/electronics9040670} {\bibfield
  {journal} {\bibinfo  {journal} {Electronics}\ }\textbf {\bibinfo {volume}
  {9}},\ \bibinfo {pages} {670} (\bibinfo {year} {2020})}\BibitemShut {NoStop}%
\bibitem [{\citenamefont {Liu}\ \emph {et~al.}(2015)\citenamefont {Liu},
  \citenamefont {Yan}, \citenamefont {Yang}, \citenamefont {Song},
  \citenamefont {Li}, \citenamefont {Liu}, \citenamefont {Chen}, \citenamefont
  {Li}, \citenamefont {Wu},\ and\ \citenamefont {Jiang}}]{liu_spiking_2015}%
  \BibitemOpen
  \bibfield  {author} {\bibinfo {author} {\bibfnamefont {C.}~\bibnamefont
  {Liu}}, \bibinfo {author} {\bibfnamefont {B.}~\bibnamefont {Yan}}, \bibinfo
  {author} {\bibfnamefont {C.}~\bibnamefont {Yang}}, \bibinfo {author}
  {\bibfnamefont {L.}~\bibnamefont {Song}}, \bibinfo {author} {\bibfnamefont
  {Z.}~\bibnamefont {Li}}, \bibinfo {author} {\bibfnamefont {B.}~\bibnamefont
  {Liu}}, \bibinfo {author} {\bibfnamefont {Y.}~\bibnamefont {Chen}}, \bibinfo
  {author} {\bibfnamefont {H.}~\bibnamefont {Li}}, \bibinfo {author}
  {\bibfnamefont {Q.}~\bibnamefont {Wu}}, \ and\ \bibinfo {author}
  {\bibfnamefont {H.}~\bibnamefont {Jiang}},\ }\bibfield  {title} {\enquote
  {\bibinfo {title} {A spiking neuromorphic design with resistive crossbar},}\
  }in\ \href@noop {} {\emph {\bibinfo {booktitle} {2015 52nd
  {{ACM}}/{{EDAC}}/{{IEEE Design Automation Conference}} ({{DAC}})}}}\
  (\bibinfo  {publisher} {{IEEE}},\ \bibinfo {year} {2015})\ pp.\ \bibinfo
  {pages} {1--6}\BibitemShut {NoStop}%
\bibitem [{\citenamefont {Li}\ \emph {et~al.}(2013)\citenamefont {Li},
  \citenamefont {Shan}, \citenamefont {Hu}, \citenamefont {Wang}, \citenamefont
  {Chen},\ and\ \citenamefont {Yang}}]{li_memristor-based_2013}%
  \BibitemOpen
  \bibfield  {author} {\bibinfo {author} {\bibfnamefont {B.}~\bibnamefont
  {Li}}, \bibinfo {author} {\bibfnamefont {Y.}~\bibnamefont {Shan}}, \bibinfo
  {author} {\bibfnamefont {M.}~\bibnamefont {Hu}}, \bibinfo {author}
  {\bibfnamefont {Y.}~\bibnamefont {Wang}}, \bibinfo {author} {\bibfnamefont
  {Y.}~\bibnamefont {Chen}}, \ and\ \bibinfo {author} {\bibfnamefont
  {H.}~\bibnamefont {Yang}},\ }\bibfield  {title} {\enquote {\bibinfo {title}
  {Memristor-based approximated computation},}\ }in\ \href {\doibase
  10.1109/ISLPED.2013.6629302} {\emph {\bibinfo {booktitle} {International
  {{Symposium}} on {{Low Power Electronics}} and {{Design}} ({{ISLPED}})}}}\
  (\bibinfo  {publisher} {{IEEE}},\ \bibinfo {year} {2013})\ pp.\ \bibinfo
  {pages} {242--247}\BibitemShut {NoStop}%
\bibitem [{\citenamefont {Demler}(2018)}]{demler_mythic_2018}%
  \BibitemOpen
  \bibfield  {author} {\bibinfo {author} {\bibfnamefont {M.}~\bibnamefont
  {Demler}},\ }\bibfield  {title} {\enquote {\bibinfo {title} {{{MYTHIC
  MULTIPLIES IN A FLASH}}},}\ }\href@noop {} {\ ,\ \bibinfo {pages} {3}
  (\bibinfo {year} {2018})}\BibitemShut {NoStop}%
\bibitem [{\citenamefont {Camsari}\ \emph {et~al.}(2017)\citenamefont
  {Camsari}, \citenamefont {Faria}, \citenamefont {Sutton},\ and\ \citenamefont
  {Datta}}]{camsari_stochastic_2017}%
  \BibitemOpen
  \bibfield  {author} {\bibinfo {author} {\bibfnamefont {K.~Y.}\ \bibnamefont
  {Camsari}}, \bibinfo {author} {\bibfnamefont {R.}~\bibnamefont {Faria}},
  \bibinfo {author} {\bibfnamefont {B.~M.}\ \bibnamefont {Sutton}}, \ and\
  \bibinfo {author} {\bibfnamefont {S.}~\bibnamefont {Datta}},\ }\bibfield
  {title} {\enquote {\bibinfo {title} {Stochastic p -{{Bits}} for {{Invertible
  Logic}}},}\ }\href {\doibase 10.1103/PhysRevX.7.031014} {\bibfield  {journal}
  {\bibinfo  {journal} {Physical Review X}\ }\textbf {\bibinfo {volume} {7}}
  (\bibinfo {year} {2017}),\ 10.1103/PhysRevX.7.031014}\BibitemShut {NoStop}%
\bibitem [{\citenamefont {Kaiser}\ \emph {et~al.}(2019)\citenamefont {Kaiser},
  \citenamefont {Rustagi}, \citenamefont {Camsari}, \citenamefont {Sun},
  \citenamefont {Datta},\ and\ \citenamefont
  {Upadhyaya}}]{kaiser_subnanosecond_2019}%
  \BibitemOpen
  \bibfield  {author} {\bibinfo {author} {\bibfnamefont {J.}~\bibnamefont
  {Kaiser}}, \bibinfo {author} {\bibfnamefont {A.}~\bibnamefont {Rustagi}},
  \bibinfo {author} {\bibfnamefont {K.~Y.}\ \bibnamefont {Camsari}}, \bibinfo
  {author} {\bibfnamefont {J.~Z.}\ \bibnamefont {Sun}}, \bibinfo {author}
  {\bibfnamefont {S.}~\bibnamefont {Datta}}, \ and\ \bibinfo {author}
  {\bibfnamefont {P.}~\bibnamefont {Upadhyaya}},\ }\bibfield  {title} {\enquote
  {\bibinfo {title} {Subnanosecond {{Fluctuations}} in {{Low}}-{{Barrier
  Nanomagnets}}},}\ }\href {\doibase 10.1103/PhysRevApplied.12.054056}
  {\bibfield  {journal} {\bibinfo  {journal} {Physical Review Applied}\
  }\textbf {\bibinfo {volume} {12}},\ \bibinfo {pages} {054056} (\bibinfo
  {year} {2019})}\BibitemShut {NoStop}%
\bibitem [{\citenamefont {Kanai}\ \emph {et~al.}(2021)\citenamefont {Kanai},
  \citenamefont {Hayakawa}, \citenamefont {Ohno},\ and\ \citenamefont
  {Fukami}}]{kanai_theory_2021}%
  \BibitemOpen
  \bibfield  {author} {\bibinfo {author} {\bibfnamefont {S.}~\bibnamefont
  {Kanai}}, \bibinfo {author} {\bibfnamefont {K.}~\bibnamefont {Hayakawa}},
  \bibinfo {author} {\bibfnamefont {H.}~\bibnamefont {Ohno}}, \ and\ \bibinfo
  {author} {\bibfnamefont {S.}~\bibnamefont {Fukami}},\ }\bibfield  {title}
  {\enquote {\bibinfo {title} {Theory of relaxation time of stochastic
  nanomagnets},}\ }\href {\doibase 10.1103/PhysRevB.103.094423} {\bibfield
  {journal} {\bibinfo  {journal} {Physical Review B}\ }\textbf {\bibinfo
  {volume} {103}},\ \bibinfo {pages} {094423} (\bibinfo {year}
  {2021})}\BibitemShut {NoStop}%
\bibitem [{\citenamefont {Safranski}\ \emph {et~al.}(2021)\citenamefont
  {Safranski}, \citenamefont {Kaiser}, \citenamefont {Trouilloud},
  \citenamefont {Hashemi}, \citenamefont {Hu},\ and\ \citenamefont
  {Sun}}]{safranski_demonstration_2021}%
  \BibitemOpen
  \bibfield  {author} {\bibinfo {author} {\bibfnamefont {C.}~\bibnamefont
  {Safranski}}, \bibinfo {author} {\bibfnamefont {J.}~\bibnamefont {Kaiser}},
  \bibinfo {author} {\bibfnamefont {P.}~\bibnamefont {Trouilloud}}, \bibinfo
  {author} {\bibfnamefont {P.}~\bibnamefont {Hashemi}}, \bibinfo {author}
  {\bibfnamefont {G.}~\bibnamefont {Hu}}, \ and\ \bibinfo {author}
  {\bibfnamefont {J.~Z.}\ \bibnamefont {Sun}},\ }\bibfield  {title} {\enquote
  {\bibinfo {title} {Demonstration of {{Nanosecond Operation}} in {{Stochastic
  Magnetic Tunnel Junctions}}},}\ }\href {\doibase
  10.1021/acs.nanolett.0c04652} {\bibfield  {journal} {\bibinfo  {journal}
  {Nano Letters}\ }\textbf {\bibinfo {volume} {21}},\ \bibinfo {pages}
  {2040--2045} (\bibinfo {year} {2021})}\BibitemShut {NoStop}%
\bibitem [{\citenamefont {Hayakawa}\ \emph {et~al.}(2021)\citenamefont
  {Hayakawa}, \citenamefont {Kanai}, \citenamefont {Funatsu}, \citenamefont
  {Igarashi}, \citenamefont {Jinnai}, \citenamefont {Borders}, \citenamefont
  {Ohno},\ and\ \citenamefont {Fukami}}]{hayakawa_nanosecond_2021}%
  \BibitemOpen
  \bibfield  {author} {\bibinfo {author} {\bibfnamefont {K.}~\bibnamefont
  {Hayakawa}}, \bibinfo {author} {\bibfnamefont {S.}~\bibnamefont {Kanai}},
  \bibinfo {author} {\bibfnamefont {T.}~\bibnamefont {Funatsu}}, \bibinfo
  {author} {\bibfnamefont {J.}~\bibnamefont {Igarashi}}, \bibinfo {author}
  {\bibfnamefont {B.}~\bibnamefont {Jinnai}}, \bibinfo {author} {\bibfnamefont
  {W.~A.}\ \bibnamefont {Borders}}, \bibinfo {author} {\bibfnamefont
  {H.}~\bibnamefont {Ohno}}, \ and\ \bibinfo {author} {\bibfnamefont
  {S.}~\bibnamefont {Fukami}},\ }\bibfield  {title} {\enquote {\bibinfo {title}
  {Nanosecond {{Random Telegraph Noise}} in {{In}}-{{Plane Magnetic Tunnel
  Junctions}}},}\ }\href {\doibase 10.1103/PhysRevLett.126.117202} {\bibfield
  {journal} {\bibinfo  {journal} {Physical Review Letters}\ }\textbf {\bibinfo
  {volume} {126}},\ \bibinfo {pages} {117202} (\bibinfo {year}
  {2021})}\BibitemShut {NoStop}%
\bibitem [{\citenamefont {Hassan}, \citenamefont {Datta},\ and\ \citenamefont
  {Camsari}(2021)}]{hassan_quantitative_2021}%
  \BibitemOpen
  \bibfield  {author} {\bibinfo {author} {\bibfnamefont {O.}~\bibnamefont
  {Hassan}}, \bibinfo {author} {\bibfnamefont {S.}~\bibnamefont {Datta}}, \
  and\ \bibinfo {author} {\bibfnamefont {K.~Y.}\ \bibnamefont {Camsari}},\
  }\bibfield  {title} {\enquote {\bibinfo {title} {Quantitative {{Evaluation}}
  of {{Hardware Binary Stochastic Neurons}}},}\ }\href {\doibase
  10.1103/PhysRevApplied.15.064046} {\bibfield  {journal} {\bibinfo  {journal}
  {Physical Review Applied}\ }\textbf {\bibinfo {volume} {15}},\ \bibinfo
  {pages} {064046} (\bibinfo {year} {2021})}\BibitemShut {NoStop}%
\bibitem [{\citenamefont {Borders}\ \emph {et~al.}(2019)\citenamefont
  {Borders}, \citenamefont {Pervaiz}, \citenamefont {Fukami}, \citenamefont
  {Camsari}, \citenamefont {Ohno},\ and\ \citenamefont
  {Datta}}]{borders_integer_2019}%
  \BibitemOpen
  \bibfield  {author} {\bibinfo {author} {\bibfnamefont {W.~A.}\ \bibnamefont
  {Borders}}, \bibinfo {author} {\bibfnamefont {A.~Z.}\ \bibnamefont
  {Pervaiz}}, \bibinfo {author} {\bibfnamefont {S.}~\bibnamefont {Fukami}},
  \bibinfo {author} {\bibfnamefont {K.~Y.}\ \bibnamefont {Camsari}}, \bibinfo
  {author} {\bibfnamefont {H.}~\bibnamefont {Ohno}}, \ and\ \bibinfo {author}
  {\bibfnamefont {S.}~\bibnamefont {Datta}},\ }\bibfield  {title} {\enquote
  {\bibinfo {title} {Integer factorization using stochastic magnetic tunnel
  junctions},}\ }\href {\doibase 10.1038/s41586-019-1557-9} {\bibfield
  {journal} {\bibinfo  {journal} {Nature}\ }\textbf {\bibinfo {volume} {573}},\
  \bibinfo {pages} {390--393} (\bibinfo {year} {2019})}\BibitemShut {NoStop}%
\bibitem [{\citenamefont {Vigna}(2017)}]{vigna_further_2017}%
  \BibitemOpen
  \bibfield  {author} {\bibinfo {author} {\bibfnamefont {S.}~\bibnamefont
  {Vigna}},\ }\bibfield  {title} {\enquote {\bibinfo {title} {Further
  scramblings of {{Marsaglia}}'s xorshift generators},}\ }\href {\doibase
  10.1016/j.cam.2016.11.006} {\bibfield  {journal} {\bibinfo  {journal}
  {Journal of Computational and Applied Mathematics}\ }\textbf {\bibinfo
  {volume} {315}},\ \bibinfo {pages} {175--181} (\bibinfo {year}
  {2017})}\BibitemShut {NoStop}%
\bibitem [{\citenamefont {Kaiser}\ \emph {et~al.}(2020)\citenamefont {Kaiser},
  \citenamefont {Faria}, \citenamefont {Camsari},\ and\ \citenamefont
  {Datta}}]{kaiser_probabilistic_2020}%
  \BibitemOpen
  \bibfield  {author} {\bibinfo {author} {\bibfnamefont {J.}~\bibnamefont
  {Kaiser}}, \bibinfo {author} {\bibfnamefont {R.}~\bibnamefont {Faria}},
  \bibinfo {author} {\bibfnamefont {K.~Y.}\ \bibnamefont {Camsari}}, \ and\
  \bibinfo {author} {\bibfnamefont {S.}~\bibnamefont {Datta}},\ }\bibfield
  {title} {\enquote {\bibinfo {title} {Probabilistic {{Circuits}} for
  {{Autonomous Learning}}: A simulation study},}\ }\href {\doibase
  10.3389/fncom.2020.00014} {\bibfield  {journal} {\bibinfo  {journal}
  {Frontiers in Computational Neuroscience}\ }\textbf {\bibinfo {volume} {14}}
  (\bibinfo {year} {2020}),\ 10.3389/fncom.2020.00014}\BibitemShut {NoStop}%
\bibitem [{\citenamefont {Kaiser}\ \emph
  {et~al.}(2021{\natexlab{b}})\citenamefont {Kaiser}, \citenamefont {Borders},
  \citenamefont {Camsari}, \citenamefont {Fukami}, \citenamefont {Ohno},\ and\
  \citenamefont {Datta}}]{kaiser_hardware-aware_2021}%
  \BibitemOpen
  \bibfield  {author} {\bibinfo {author} {\bibfnamefont {J.}~\bibnamefont
  {Kaiser}}, \bibinfo {author} {\bibfnamefont {W.~A.}\ \bibnamefont {Borders}},
  \bibinfo {author} {\bibfnamefont {K.~Y.}\ \bibnamefont {Camsari}}, \bibinfo
  {author} {\bibfnamefont {S.}~\bibnamefont {Fukami}}, \bibinfo {author}
  {\bibfnamefont {H.}~\bibnamefont {Ohno}}, \ and\ \bibinfo {author}
  {\bibfnamefont {S.}~\bibnamefont {Datta}},\ }\bibfield  {title} {\enquote
  {\bibinfo {title} {Hardware-aware in-situ {{Boltzmann}} machine learning
  using stochastic magnetic tunnel junctions},}\ }\href@noop {} {\bibfield
  {journal} {\bibinfo  {journal} {arXiv:2102.05137 [cond-mat]}\ } (\bibinfo
  {year} {2021}{\natexlab{b}})},\ \Eprint {http://arxiv.org/abs/2102.05137}
  {arXiv:2102.05137 [cond-mat]} \BibitemShut {NoStop}%
\bibitem [{\citenamefont {Bishop}(2013)}]{bishop_pattern_2013}%
  \BibitemOpen
  \bibfield  {author} {\bibinfo {author} {\bibfnamefont {C.~M.}\ \bibnamefont
  {Bishop}},\ }\href@noop {} {\emph {\bibinfo {title} {Pattern {{Recognition}}
  and {{Machine Learning}}: All "Just the {{Facts}} 101" {{Material}}}}}\
  (\bibinfo  {publisher} {{Springer (India) Private Limited}},\ \bibinfo {year}
  {2013})\BibitemShut {NoStop}%
\bibitem [{\citenamefont {Faria}\ \emph {et~al.}(2021)\citenamefont {Faria},
  \citenamefont {Kaiser}, \citenamefont {Camsari},\ and\ \citenamefont
  {Datta}}]{faria_hardware_2021}%
  \BibitemOpen
  \bibfield  {author} {\bibinfo {author} {\bibfnamefont {R.}~\bibnamefont
  {Faria}}, \bibinfo {author} {\bibfnamefont {J.}~\bibnamefont {Kaiser}},
  \bibinfo {author} {\bibfnamefont {K.~Y.}\ \bibnamefont {Camsari}}, \ and\
  \bibinfo {author} {\bibfnamefont {S.}~\bibnamefont {Datta}},\ }\bibfield
  {title} {\enquote {\bibinfo {title} {Hardware {{Design}} for {{Autonomous
  Bayesian Networks}}},}\ }\href {\doibase 10.3389/fncom.2021.584797}
  {\bibfield  {journal} {\bibinfo  {journal} {Frontiers in Computational
  Neuroscience}\ }\textbf {\bibinfo {volume} {15}} (\bibinfo {year} {2021}),\
  10.3389/fncom.2021.584797}\BibitemShut {NoStop}%
\bibitem [{\citenamefont {Faria}, \citenamefont {Camsari},\ and\ \citenamefont
  {Datta}(2018)}]{faria_implementing_2018}%
  \BibitemOpen
  \bibfield  {author} {\bibinfo {author} {\bibfnamefont {R.}~\bibnamefont
  {Faria}}, \bibinfo {author} {\bibfnamefont {K.~Y.}\ \bibnamefont {Camsari}},
  \ and\ \bibinfo {author} {\bibfnamefont {S.}~\bibnamefont {Datta}},\
  }\bibfield  {title} {\enquote {\bibinfo {title} {Implementing {{Bayesian}}
  networks with embedded stochastic {{MRAM}}},}\ }\href {\doibase
  10.1063/1.5021332} {\bibfield  {journal} {\bibinfo  {journal} {AIP Advances}\
  }\textbf {\bibinfo {volume} {8}},\ \bibinfo {pages} {045101} (\bibinfo {year}
  {2018})}\BibitemShut {NoStop}%
\bibitem [{\citenamefont {Koller}\ and\ \citenamefont
  {Friedman}(2009)}]{koller_probabilistic_2009}%
  \BibitemOpen
  \bibfield  {author} {\bibinfo {author} {\bibfnamefont {D.}~\bibnamefont
  {Koller}}\ and\ \bibinfo {author} {\bibfnamefont {N.}~\bibnamefont
  {Friedman}},\ }\href@noop {} {\emph {\bibinfo {title} {Probabilistic
  {{Graphical Models}}: Principles and {{Techniques}}}}}\ (\bibinfo
  {publisher} {{MIT Press}},\ \bibinfo {year} {2009})\BibitemShut {NoStop}%
\bibitem [{\citenamefont {{Behin-Aein}}, \citenamefont {Diep},\ and\
  \citenamefont {Datta}(2016)}]{behin-aein_building_2016-2}%
  \BibitemOpen
  \bibfield  {author} {\bibinfo {author} {\bibfnamefont {B.}~\bibnamefont
  {{Behin-Aein}}}, \bibinfo {author} {\bibfnamefont {V.}~\bibnamefont {Diep}},
  \ and\ \bibinfo {author} {\bibfnamefont {S.}~\bibnamefont {Datta}},\
  }\bibfield  {title} {\enquote {\bibinfo {title} {A building block for
  hardware belief networks},}\ }\href {\doibase 10.1038/srep29893} {\bibfield
  {journal} {\bibinfo  {journal} {Scientific Reports}\ }\textbf {\bibinfo
  {volume} {6}},\ \bibinfo {pages} {29893} (\bibinfo {year}
  {2016})}\BibitemShut {NoStop}%
\bibitem [{\citenamefont {Yang}\ \emph {et~al.}(2020)\citenamefont {Yang},
  \citenamefont {Malhotra}, \citenamefont {Lu},\ and\ \citenamefont
  {Sengupta}}]{yang_all-spin_2020}%
  \BibitemOpen
  \bibfield  {author} {\bibinfo {author} {\bibfnamefont {K.}~\bibnamefont
  {Yang}}, \bibinfo {author} {\bibfnamefont {A.}~\bibnamefont {Malhotra}},
  \bibinfo {author} {\bibfnamefont {S.}~\bibnamefont {Lu}}, \ and\ \bibinfo
  {author} {\bibfnamefont {A.}~\bibnamefont {Sengupta}},\ }\bibfield  {title}
  {\enquote {\bibinfo {title} {All-{{Spin Bayesian Neural Networks}}},}\ }\href
  {\doibase 10.1109/TED.2020.2968223} {\bibfield  {journal} {\bibinfo
  {journal} {IEEE Transactions on Electron Devices}\ }\textbf {\bibinfo
  {volume} {67}},\ \bibinfo {pages} {1340--1347} (\bibinfo {year}
  {2020})}\BibitemShut {NoStop}%
\bibitem [{\citenamefont {Martello}, \citenamefont {Pisinger},\ and\
  \citenamefont {Toth}(1999)}]{martello_dynamic_1999}%
  \BibitemOpen
  \bibfield  {author} {\bibinfo {author} {\bibfnamefont {S.}~\bibnamefont
  {Martello}}, \bibinfo {author} {\bibfnamefont {D.}~\bibnamefont {Pisinger}},
  \ and\ \bibinfo {author} {\bibfnamefont {P.}~\bibnamefont {Toth}},\
  }\bibfield  {title} {\enquote {\bibinfo {title} {Dynamic {{Programming}} and
  {{Strong Bounds}} for the 0-1 {{Knapsack Problem}}},}\ }\href@noop {}
  {\bibfield  {journal} {\bibinfo  {journal} {Management Science}\ }\textbf
  {\bibinfo {volume} {45}},\ \bibinfo {pages} {414--424} (\bibinfo {year}
  {1999})}\BibitemShut {NoStop}%
\bibitem [{\citenamefont {Zhang}\ \emph {et~al.}(2021)\citenamefont {Zhang},
  \citenamefont {Bashizade}, \citenamefont {Wang}, \citenamefont {Mukherjee},\
  and\ \citenamefont {Lebeck}}]{zhang_statistical_2021}%
  \BibitemOpen
  \bibfield  {author} {\bibinfo {author} {\bibfnamefont {X.}~\bibnamefont
  {Zhang}}, \bibinfo {author} {\bibfnamefont {R.}~\bibnamefont {Bashizade}},
  \bibinfo {author} {\bibfnamefont {Y.}~\bibnamefont {Wang}}, \bibinfo {author}
  {\bibfnamefont {S.}~\bibnamefont {Mukherjee}}, \ and\ \bibinfo {author}
  {\bibfnamefont {A.~R.}\ \bibnamefont {Lebeck}},\ }\bibfield  {title}
  {\enquote {\bibinfo {title} {Statistical robustness of {{Markov}} chain
  {{Monte Carlo}} accelerators},}\ }in\ \href {\doibase
  10.1145/3445814.3446697} {\emph {\bibinfo {booktitle} {Proceedings of the
  26th {{ACM International Conference}} on {{Architectural Support}} for
  {{Programming Languages}} and {{Operating Systems}}}}}\ (\bibinfo
  {publisher} {{ACM}},\ \bibinfo {address} {{Virtual USA}},\ \bibinfo {year}
  {2021})\ pp.\ \bibinfo {pages} {959--974}\BibitemShut {NoStop}%
\bibitem [{\citenamefont {Liu}, \citenamefont {Liang},\ and\ \citenamefont
  {Wong}(2000)}]{liu_multiple-try_2000}%
  \BibitemOpen
  \bibfield  {author} {\bibinfo {author} {\bibfnamefont {J.~S.}\ \bibnamefont
  {Liu}}, \bibinfo {author} {\bibfnamefont {F.}~\bibnamefont {Liang}}, \ and\
  \bibinfo {author} {\bibfnamefont {W.~H.}\ \bibnamefont {Wong}},\ }\bibfield
  {title} {\enquote {\bibinfo {title} {The {{Multiple}}-{{Try Method}} and
  {{Local Optimization}} in {{Metropolis Sampling}}},}\ }\href {\doibase
  10.1080/01621459.2000.10473908} {\bibfield  {journal} {\bibinfo  {journal}
  {Journal of the American Statistical Association}\ }\textbf {\bibinfo
  {volume} {95}},\ \bibinfo {pages} {121--134} (\bibinfo {year}
  {2000})}\BibitemShut {NoStop}%
\bibitem [{\citenamefont {Kellerer}, \citenamefont {Pferschy},\ and\
  \citenamefont {Pisinger}(2004)}]{kellerer_knapsack_2004}%
  \BibitemOpen
  \bibfield  {author} {\bibinfo {author} {\bibfnamefont {H.}~\bibnamefont
  {Kellerer}}, \bibinfo {author} {\bibfnamefont {U.}~\bibnamefont {Pferschy}},
  \ and\ \bibinfo {author} {\bibfnamefont {D.}~\bibnamefont {Pisinger}},\
  }\href {\doibase 10.1007/978-3-540-24777-7} {\emph {\bibinfo {title}
  {Knapsack {{Problems}}}}}\ (\bibinfo {year} {2004})\BibitemShut {NoStop}%
\bibitem [{\citenamefont {Geman}\ and\ \citenamefont
  {Geman}(1984)}]{geman_stochastic_1984-2}%
  \BibitemOpen
  \bibfield  {author} {\bibinfo {author} {\bibfnamefont {S.}~\bibnamefont
  {Geman}}\ and\ \bibinfo {author} {\bibfnamefont {D.}~\bibnamefont {Geman}},\
  }\bibfield  {title} {\enquote {\bibinfo {title} {Stochastic {{Relaxation}},
  {{Gibbs Distributions}}, and the {{Bayesian Restoration}} of {{Images}}},}\
  }\href {\doibase 10.1109/TPAMI.1984.4767596} {\bibfield  {journal} {\bibinfo
  {journal} {IEEE Transactions on Pattern Analysis and Machine Intelligence}\
  }\textbf {\bibinfo {volume} {PAMI-6}},\ \bibinfo {pages} {721--741} (\bibinfo
  {year} {1984})}\BibitemShut {NoStop}%
\bibitem [{\citenamefont {Hastings}(1970)}]{hastings_monte_1970}%
  \BibitemOpen
  \bibfield  {author} {\bibinfo {author} {\bibfnamefont {W.~K.}\ \bibnamefont
  {Hastings}},\ }\bibfield  {title} {\enquote {\bibinfo {title} {Monte
  {{Carlo}} sampling methods using {{Markov}} chains and their applications},}\
  }\href {\doibase 10.1093/biomet/57.1.97} {\bibfield  {journal} {\bibinfo
  {journal} {Biometrika}\ }\textbf {\bibinfo {volume} {57}},\ \bibinfo {pages}
  {97--109} (\bibinfo {year} {1970})}\BibitemShut {NoStop}%
\bibitem [{\citenamefont {Lv}, \citenamefont {Bloom},\ and\ \citenamefont
  {Wang}(2019)}]{lv_experimental_2019}%
  \BibitemOpen
  \bibfield  {author} {\bibinfo {author} {\bibfnamefont {Y.}~\bibnamefont
  {Lv}}, \bibinfo {author} {\bibfnamefont {R.~P.}\ \bibnamefont {Bloom}}, \
  and\ \bibinfo {author} {\bibfnamefont {J.-P.}\ \bibnamefont {Wang}},\
  }\bibfield  {title} {\enquote {\bibinfo {title} {Experimental
  {{Demonstration}} of {{Probabilistic Spin Logic}} by {{Magnetic Tunnel
  Junctions}}},}\ }\href {\doibase 10.1109/LMAG.2019.2957258} {\bibfield
  {journal} {\bibinfo  {journal} {IEEE Magnetics Letters}\ }\textbf {\bibinfo
  {volume} {10}},\ \bibinfo {pages} {1--5} (\bibinfo {year}
  {2019})}\BibitemShut {NoStop}%
\bibitem [{\citenamefont {Aadit}\ \emph {et~al.}(2021)\citenamefont {Aadit},
  \citenamefont {Grimaldi}, \citenamefont {Carpentieri}, \citenamefont
  {Theogarajan}, \citenamefont {Finocchio},\ and\ \citenamefont
  {Camsari}}]{aadit_computing_2021}%
  \BibitemOpen
  \bibfield  {author} {\bibinfo {author} {\bibfnamefont {N.~A.}\ \bibnamefont
  {Aadit}}, \bibinfo {author} {\bibfnamefont {A.}~\bibnamefont {Grimaldi}},
  \bibinfo {author} {\bibfnamefont {M.}~\bibnamefont {Carpentieri}}, \bibinfo
  {author} {\bibfnamefont {L.}~\bibnamefont {Theogarajan}}, \bibinfo {author}
  {\bibfnamefont {G.}~\bibnamefont {Finocchio}}, \ and\ \bibinfo {author}
  {\bibfnamefont {K.~Y.}\ \bibnamefont {Camsari}},\ }\bibfield  {title}
  {\enquote {\bibinfo {title} {Computing with {{Invertible Logic}}:
  Combinatorial {{Optimization}} with {{Probabilistic Bits}}},}\ }\href@noop {}
  {\bibfield  {journal} {\bibinfo  {journal} {IEEE International Electron
  Devices Meeting (IEDM) (To appear)}\ } (\bibinfo {year} {2021})}\BibitemShut
  {NoStop}%
\bibitem [{\citenamefont {Chowdhury}, \citenamefont {Camsari},\ and\
  \citenamefont {Datta}(2020)}]{chowdhury_emulating_2020}%
  \BibitemOpen
  \bibfield  {author} {\bibinfo {author} {\bibfnamefont {S.}~\bibnamefont
  {Chowdhury}}, \bibinfo {author} {\bibfnamefont {K.~Y.}\ \bibnamefont
  {Camsari}}, \ and\ \bibinfo {author} {\bibfnamefont {S.}~\bibnamefont
  {Datta}},\ }\bibfield  {title} {\enquote {\bibinfo {title} {Emulating
  {{Quantum Interference}} with {{Generalized Ising Machines}}},}\ }\href@noop
  {} {\bibfield  {journal} {\bibinfo  {journal} {arXiv:2007.07379 [cond-mat,
  physics:quant-ph]}\ } (\bibinfo {year} {2020})},\ \Eprint
  {http://arxiv.org/abs/2007.07379} {arXiv:2007.07379 [cond-mat,
  physics:quant-ph]} \BibitemShut {NoStop}%
\bibitem [{\citenamefont {Camsari}, \citenamefont {Chowdhury},\ and\
  \citenamefont {Datta}(2019)}]{camsari_scalable_2019}%
  \BibitemOpen
  \bibfield  {author} {\bibinfo {author} {\bibfnamefont {K.~Y.}\ \bibnamefont
  {Camsari}}, \bibinfo {author} {\bibfnamefont {S.}~\bibnamefont {Chowdhury}},
  \ and\ \bibinfo {author} {\bibfnamefont {S.}~\bibnamefont {Datta}},\
  }\bibfield  {title} {\enquote {\bibinfo {title} {Scalable {{Emulation}} of
  {{Sign}}-{{Problem}}\textendash{{Free Hamiltonians}} with
  {{Room}}-{{Temperature}} \$p\$-bits},}\ }\href {\doibase
  10.1103/PhysRevApplied.12.034061} {\bibfield  {journal} {\bibinfo  {journal}
  {Phys. Rev. Applied}\ }\textbf {\bibinfo {volume} {12}},\ \bibinfo {pages}
  {034061} (\bibinfo {year} {2019})}\BibitemShut {NoStop}%
\bibitem [{\citenamefont {Bravyi}\ \emph {et~al.}(2008)\citenamefont {Bravyi},
  \citenamefont {Divincenzo}, \citenamefont {Oliveira},\ and\ \citenamefont
  {Terhal}}]{bravyi_complexity_2008}%
  \BibitemOpen
  \bibfield  {author} {\bibinfo {author} {\bibfnamefont {S.}~\bibnamefont
  {Bravyi}}, \bibinfo {author} {\bibfnamefont {D.~P.}\ \bibnamefont
  {Divincenzo}}, \bibinfo {author} {\bibfnamefont {R.}~\bibnamefont
  {Oliveira}}, \ and\ \bibinfo {author} {\bibfnamefont {B.~M.}\ \bibnamefont
  {Terhal}},\ }\bibfield  {title} {\enquote {\bibinfo {title} {The complexity
  of stoquastic local {{Hamiltonian}} problems},}\ }\href@noop {} {\bibfield
  {journal} {\bibinfo  {journal} {Quantum Information \& Computation}\ }\textbf
  {\bibinfo {volume} {8}},\ \bibinfo {pages} {361--385} (\bibinfo {year}
  {2008})}\BibitemShut {NoStop}%
\bibitem [{\citenamefont {Drineas}, \citenamefont {Kannan},\ and\ \citenamefont
  {Mahoney}(2006{\natexlab{a}})}]{drineas_fast_2006-1}%
  \BibitemOpen
  \bibfield  {author} {\bibinfo {author} {\bibfnamefont {P.}~\bibnamefont
  {Drineas}}, \bibinfo {author} {\bibfnamefont {R.}~\bibnamefont {Kannan}}, \
  and\ \bibinfo {author} {\bibfnamefont {M.~W.}\ \bibnamefont {Mahoney}},\
  }\bibfield  {title} {\enquote {\bibinfo {title} {Fast {{Monte Carlo
  Algorithms}} for {{Matrices I}}: Approximating {{Matrix Multiplication}}},}\
  }\href {\doibase 10.1137/S0097539704442684} {\bibfield  {journal} {\bibinfo
  {journal} {SIAM Journal on Computing}\ }\textbf {\bibinfo {volume} {36}},\
  \bibinfo {pages} {132--157} (\bibinfo {year}
  {2006}{\natexlab{a}})}\BibitemShut {NoStop}%
\bibitem [{\citenamefont {Drineas}, \citenamefont {Kannan},\ and\ \citenamefont
  {Mahoney}(2006{\natexlab{b}})}]{drineas_fast_2006-2}%
  \BibitemOpen
  \bibfield  {author} {\bibinfo {author} {\bibfnamefont {P.}~\bibnamefont
  {Drineas}}, \bibinfo {author} {\bibfnamefont {R.}~\bibnamefont {Kannan}}, \
  and\ \bibinfo {author} {\bibfnamefont {M.~W.}\ \bibnamefont {Mahoney}},\
  }\bibfield  {title} {\enquote {\bibinfo {title} {Fast {{Monte Carlo
  Algorithms}} for {{Matrices II}}: Computing a {{Low}}-{{Rank Approximation}}
  to a {{Matrix}}},}\ }\href {\doibase 10.1137/S0097539704442696} {\bibfield
  {journal} {\bibinfo  {journal} {SIAM Journal on Computing}\ }\textbf
  {\bibinfo {volume} {36}},\ \bibinfo {pages} {158--183} (\bibinfo {year}
  {2006}{\natexlab{b}})}\BibitemShut {NoStop}%
\bibitem [{\citenamefont {Buluc}\ \emph {et~al.}(2021)\citenamefont {Buluc},
  \citenamefont {Kolda}, \citenamefont {Wild}, \citenamefont {Anitescu},
  \citenamefont {DeGennaro}, \citenamefont {Jakeman}, \citenamefont {Kamath},
  \citenamefont {Ramakrishnan}, \citenamefont {Kannan}, \citenamefont {Lopes},
  \citenamefont {Martinsson}, \citenamefont {Myers}, \citenamefont {Nelson},
  \citenamefont {Restrepo}, \citenamefont {Seshadhri}, \citenamefont {Vrabie},
  \citenamefont {Wohlberg}, \citenamefont {Wright}, \citenamefont {Yang},\ and\
  \citenamefont {Zwart}}]{buluc_randomized_2021}%
  \BibitemOpen
  \bibfield  {author} {\bibinfo {author} {\bibfnamefont {A.}~\bibnamefont
  {Buluc}}, \bibinfo {author} {\bibfnamefont {T.~G.}\ \bibnamefont {Kolda}},
  \bibinfo {author} {\bibfnamefont {S.~M.}\ \bibnamefont {Wild}}, \bibinfo
  {author} {\bibfnamefont {M.}~\bibnamefont {Anitescu}}, \bibinfo {author}
  {\bibfnamefont {A.}~\bibnamefont {DeGennaro}}, \bibinfo {author}
  {\bibfnamefont {J.}~\bibnamefont {Jakeman}}, \bibinfo {author} {\bibfnamefont
  {C.}~\bibnamefont {Kamath}}, \bibinfo {author} {\bibnamefont {Ramakrishnan}},
  \bibinfo {author} {\bibnamefont {Kannan}}, \bibinfo {author} {\bibfnamefont
  {M.~E.}\ \bibnamefont {Lopes}}, \bibinfo {author} {\bibfnamefont {P.-G.}\
  \bibnamefont {Martinsson}}, \bibinfo {author} {\bibfnamefont
  {K.}~\bibnamefont {Myers}}, \bibinfo {author} {\bibfnamefont
  {J.}~\bibnamefont {Nelson}}, \bibinfo {author} {\bibfnamefont {J.~M.}\
  \bibnamefont {Restrepo}}, \bibinfo {author} {\bibfnamefont {C.}~\bibnamefont
  {Seshadhri}}, \bibinfo {author} {\bibfnamefont {D.}~\bibnamefont {Vrabie}},
  \bibinfo {author} {\bibfnamefont {B.}~\bibnamefont {Wohlberg}}, \bibinfo
  {author} {\bibfnamefont {S.~J.}\ \bibnamefont {Wright}}, \bibinfo {author}
  {\bibfnamefont {C.}~\bibnamefont {Yang}}, \ and\ \bibinfo {author}
  {\bibfnamefont {P.}~\bibnamefont {Zwart}},\ }\bibfield  {title} {\enquote
  {\bibinfo {title} {Randomized {{Algorithms}} for {{Scientific Computing}}
  ({{RASC}})},}\ }\href@noop {} {\bibfield  {journal} {\bibinfo  {journal}
  {arXiv:2104.11079 [cs]}\ } (\bibinfo {year} {2021})},\ \Eprint
  {http://arxiv.org/abs/2104.11079} {arXiv:2104.11079 [cs]} \BibitemShut
  {NoStop}%
\end{thebibliography}%

\end{document}